\documentclass[a4paper,11pt]{article}
\usepackage{jcappub} 

\usepackage[nameinlink,noabbrev]{cleveref}
\usepackage{xspace}
\usepackage{url}
\usepackage{aas_macros}
\usepackage{scalerel}
\usepackage{booktabs}
\usepackage{orcidlink}

\title{
Constraining GREA, an alternative
theory accounting for the present cosmic acceleration
}

\newcommand{\fde}{\ensuremath{f_{\rm DE}}}

\newcommand{\chisq}{\ensuremath{\chi^2}}
\newcommand{\dchisq}{\ensuremath{\Delta\chi^2}}

\newcommand{\lcdm}{$\Lambda$CDM}
\newcommand{\wcdm}{$w$CDM}
\newcommand{\wowacdm}{$w_0w_a$CDM}

\newcommand{\Omo}{\ensuremath{\Omega_{\text{m},0}}}

\newcommand{\zdrag}{\ensuremath{z_\text{drag}}}

\newcommand{\cobaya}{\textsc{cobaya}}
\newcommand{\gd}{\texttt{GetDist}}
\newcommand{\kmsMpc}{\,{\rm km\,s^{-1}\,Mpc^{-1}}}

\newcommand{\panp}{Pantheon$\texttt{+}$}
\newcommand{\shoes}{SH$_0$ES }

\crefname{equation}{Eq.}{Eqs.}
\crefname{section}{Section}{Sections}
\crefname{figure}{Fig.}{Figs.}
\crefname{table}{Table}{Tables}
\crefname{appendix}{Appendix}{Appendices}
\Crefname{figure}{Figure}{Figures}
\Crefname{equation}{Equation}{Equations}
\Crefname{section}{Section}{Sections}
\Crefname{table}{Table}{Tables}

\definecolor{llgray}{gray}{0.93}
\definecolor{lgray}{gray}{0.83}
\definecolor{deepmagenta}{rgb}{0.8, 0.0, 0.8}
\definecolor{ballblue}{rgb}{0.13, 0.67, 0.8}
\definecolor{celestialblue}{rgb}{0.29, 0.59, 0.82}
\definecolor{RedWine}{rgb}{0.743,0,0}
\definecolor{DarkGreen}{rgb}{0,0.6,0}

\affiliation{Affiliations are in Appendix \ref{sec:affiliations}}
\emailAdd{calderon@fzu.cz}
\emailAdd{juan.garciabellido@uam.es}

\author[a]{{R.~Calderon}\orcidlink{0000-0002-8215-7292},}
\author[b]{{J.~Garc\'ia-Bellido}\orcidlink{0000-0002-9370-8360},}
\author[c]{{B.~Vos-Ginés}\orcidlink{0000-0002-1803-1169},}
\author[c]{{V.~Gonzalez-Perez}\orcidlink{0000-0001-9938-2755},}
\author[d,e]{{A.~Shafieloo}\orcidlink{0000-0001-6815-0337},}
\author[f]{{J.~Aguilar},}
\author[g]{{S.~Ahlen}\orcidlink{0000-0001-6098-7247},}
\author[h,i]{{D.~Bianchi}\orcidlink{0000-0001-9712-0006},}
\author[j]{{D.~Brooks},}
\author[f]{{T.~Claybaugh},}
\author[k]{{A.~de la Macorra}\orcidlink{0000-0002-1769-1640},}
\author[l,m]{{J.~E.~Forero-Romero}\orcidlink{0000-0002-2890-3725},}
\author[n,o,p]{{E.~Gaztañaga}\orcidlink{0000-0001-9632-0815},}
\author[f,q]{{S.~Gontcho A Gontcho}\orcidlink{0000-0003-3142-233X},}
\author[r]{{G.~Gutierrez},}
\author[s,t,u]{{K.~Honscheid}\orcidlink{0000-0002-6550-2023},}
\author[v]{{C.~Howlett}\orcidlink{0000-0002-1081-9410},}
\author[w]{{M.~Ishak}\orcidlink{0000-0002-6024-466X},}
\author[x]{{R.~Joyce}\orcidlink{0000-0003-0201-5241},}
\author[y]{{R.~Kehoe},}
\author[f]{{T.~Kisner}\orcidlink{0000-0003-3510-7134},}
\author[f]{{A.~Kremin}\orcidlink{0000-0001-6356-7424},}
\author[j]{{O.~Lahav},}
\author[f]{{A.~Lambert},}
\author[f]{{M.~Landriau}\orcidlink{0000-0003-1838-8528},}
\author[z,aa]{{M.~Manera}\orcidlink{0000-0003-4962-8934},}
\author[ab,aa]{{R.~Miquel},}
\author[ac]{{F.~Prada}\orcidlink{0000-0001-7145-8674},}
\author[ad]{{I.~P\'erez-R\`afols}\orcidlink{0000-0001-6979-0125},}
\author[ae]{{E.~Sanchez}\orcidlink{0000-0002-9646-8198},}
\author[f]{{D.~Schlegel},}
\author[af,ag]{{M.~Schubnell},}
\author[f]{{J.~Silber}\orcidlink{0000-0002-3461-0320},}
\author[x]{{D.~Sprayberry},}
\author[ag]{{G.~Tarl\'{e}}\orcidlink{0000-0003-1704-0781},}
\author[x]{{B.~A.~Weaver},}
\author[ah]{{H.~Zou}\orcidlink{0000-0002-6684-3997}}

\abstract{The origin of the Universe's late-time accelerated expansion remains unknown. The General Relativistic Entropic Acceleration (GREA) theory offers a compelling alternative to $\Lambda$CDM, attributing cosmic acceleration to entropy growth associated with cosmic and black hole horizons, without invoking a cosmological constant.
We test GREA against the latest DESI DR2 Baryon Acoustic Oscillations (BAO), multiple Type Ia supernova compilations (Union3, Pantheon$\texttt{+}$, DES-SN5YR), and cosmic microwave background (CMB) distance measurements. 
While GREA is not nested within $\Lambda$CDM, it achieves a comparable goodness-of-fit, highlighting its potential as a theoretically motivated framework that circumvents some of the fine-tuning issues of the standard $\Lambda$CDM cosmology.
We find that the best-fit model features a transient phantom crossing at $z \lesssim 2$, with $w_a\equiv \mathrm{d} w(a=1)/\mathrm{d}a \simeq-0.3$, in good agreement with observations. However, its present-day value $w_0\equiv w(z=0)$ is tightly constrained to be $w_0\simeq-1$. 
Upcoming low-redshift (i.e. $z < 1$) cosmological probes, from both background and perturbations, will offer promising avenues for further exploring the viability of the GREA theory.
}

\begin{document}
\maketitle
\flushbottom

\section{Introduction}

The cause of the Universe’s accelerated expansion remains unknown. For the past two decades, the standard cosmological model ($\Lambda$CDM) has explained it through a cosmological constant, $\Lambda$ \cite{Sahni:1999gb}. While this model has been in remarkable agreement with a broad range of observations ~\cite{SupernovaSearchTeam:1998fmf,SupernovaCosmologyProject:1998vns,Planck1:2020,Alam_2021,Zhao_2022}, it faces serious theoretical challenges, such as the fine-tuning and coincidence problems \cite{Weinberg:1988cp,Carroll:2000fy}, and lacks a fundamental explanation for its two main components: dark energy and dark matter. In addition, growing observational tensions and anomalies~\cite{Bull:2015stt,Bullock:2017xww,Perivolaropoulos:2021jda,Abdalla:2022yfr,Calderon:2023obf}\textemdash including discrepancies in the inferred values of the Hubble constant~\cite{Riess:2021jrx,DiValentino:2021izs,Schoneberg:2021qvd} and the amplitude of matter fluctuations\footnote{Although the latest results from the KiDS legacy are fully compatible with the inferred values from the CMB~\cite{Wright:2025xka}.}~\cite{HSC:2018mrq,KiDS:2020suj,DES:2021wwk,DiValentino:2020vvd}\textemdash have raised questions about the validity of $\Lambda$CDM. More recently, data from the Dark Energy Spectroscopic Instrument (DESI), combined with other cosmological probes, have added to this picture by hinting at $\sim4\sigma$ deviations from a $\Lambda$CDM expansion history \cite{DESI:2024mwx,DESI:2025fii}. 
If confirmed, these results could represent the first evidence for new physics beyond $\Lambda$CDM. While they are often interpreted as signs of a dynamical dark energy component \cite{DESI:2024mwx,DESI:2024aqx,DESI2024.VII.KP7B,DESI:2024kob,Ishak:2024jhs,Y3.cpe-s1.Lodha.2025,DESI:2025wyn}, alternative explanations have been proposed, including exotic neutrino masses \cite{Elbers:2024sha,DESI:2025ejh} or non-standard dark matter evolution \cite{Chen:2025wwn}, among others \cite{Lynch:2024hzh,DESI:2025ffm,Chaussidon:2025npr,Chen:2025mlf,Mirpoorian:2025rfp,Sailer:2025lxj}.

Common approaches for modeling dark energy often rely on arbitrary parametrizations of the equation of state parameter, $w(z)=P/\rho$, or non-parametric reconstruction techniques \cite{Shafieloo:2005nd,2010PhRvD..82j3502H,2012PhRvD..85l3530S,Nesseris_2012,Calderon:2022cfj,Calderon2023}, which provide limited physical insight. Parametric models, such as $w_0w_a\rm CDM$ \cite{Chevallier:2000qy,Linder:2002et}, are widely used due to their simplicity and connection with known physics \cite{Linder:2002et,Caldwell:2005tm,dePutter:2008wt,DESI:2024kob}.
However, they risk biasing the inference if the true $w(z)$ deviates significantly from the assumed (parametric) form. Non-parametric methods offer greater flexibility, but require careful validation and can be difficult to interpret physically. A more informative strategy is to explore models grounded in theoretical principles. In this work, we investigate General Relativistic Entropic Acceleration (GREA) \cite{Garcia-Bellido:2021idr}, a theory in which cosmic acceleration arises from the entropy growth associated with cosmological horizons~\cite{Espinosa-Portales:2021cac}. GREA offers a novel, conceptually motivated alternative to $\Lambda$CDM and allows for direct connections between cosmology and fundamental physics. In the cosmological context, the GREA theory predicts that the present acceleration of the universe could arise from the growth of entropy associated with cosmic~\cite{Garcia-Bellido:2021idr} and black hole horizons~\cite{Garcia-Bellido:2024tip}, without the need to introduce a cosmological constant, providing an alternative dynamics to that of $\Lambda$CDM. The GREA theory gives a covariant formalism for out of equilibrium dynamics in the context of general relativity. A consequence of GREA is the explicit breaking of the time reversal invariance when there is entropy production. This drives an entropic force that behaves effectively like bulk viscosity with a negative effective pressure~\cite{Garcia-Bellido:2024tip}. 

Already before the recent evidence of significant deviations from the standard model of cosmology through state-of-the-art observations by DES-Y5 SNe~\cite{DES:2024tys} and DESI~\cite{DESI:2024mwx,DESI:2024aqx,DESI:2024kob,Ishak:2024jhs,DESI:2025fii,DESI:2024jxi}, GREA was found to provide an equally good description of cosmological observations than $\Lambda$CDM~\cite{Arjona:2021uxs}.
Furthermore, GREA not only explains the coincidence problem, but can also help ease the Hubble tension by shifting the coasting point (when the universe transitioned from matter domination to acceleration) to higher redshifts and extending the period of acceleration \cite{Garcia-Bellido:2024qau}. 
In a recent paper~\cite{Garcia-Bellido:2024qau}, we performed a detailed calculation of the background and linear matter perturbations with the assumption that the entropy growth responsible for the accelerated expansion comes from the homogeneous cosmological horizon.
We could show that, in this case, the dynamics can be described in terms of just one parameter, $\alpha$, the ratio of the spatial curvature to the horizon distance.
A given value of $\alpha$ defines a unique expansion history of the universe, with a dynamics that differs from $\Lambda$CDM and gives an explicit prediction for all background observables, like the values of the matter and dark energy content of the universe, the rate of expansion today, the coasting point, the value of the equation of state of dark energy and its derivative, the angular diameter and luminosity distances, etc. In this work, we use different observational datasets, including the latest baryon acoustic oscillation (BAO) measurements from DESI, to constrain the expansion history, as predicted by GREA. 

This paper is structured as follows. In \cref{sec:GREA}, we introduce the theoretical background and details of GREA. We describe the observational datasets in \cref{sec:DataMethods}, and the methodology used in the analysis in \cref{sec:methods}. Finally, we present our results in \cref{sec:results} and conclude in \cref{sec:conclusions}.

\clearpage
\section{Theoretical background}\label{sec:GREA}

\begin{figure}[h]
    \centering
    \includegraphics[width=\textwidth]{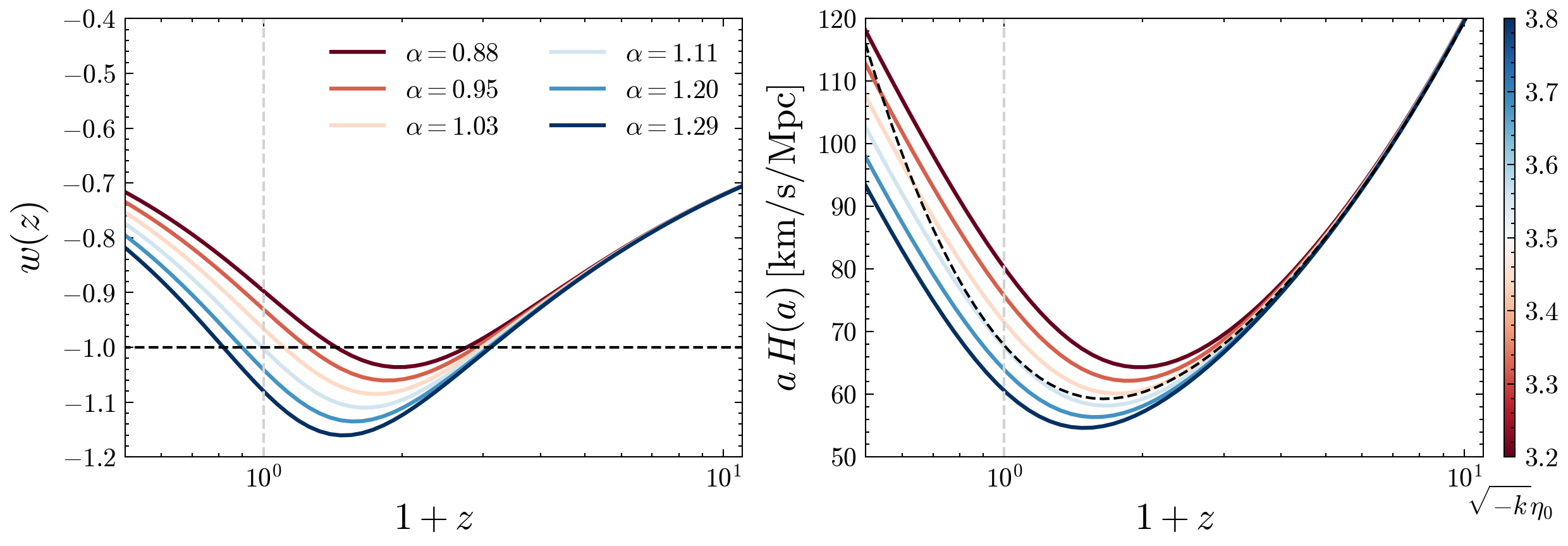}
    \caption{{\em Left}: The equation of state in GREA, for various values of the parameter $\alpha$ (or, equivalently $\sqrt{-k}\,\eta_0$), as obtained by integrating \cref{eq:tau} with initial (CMB-asymptotic) $\Omega_m=0.31,h=0.678$. Note that for some parameter combinations, the \textit{effective} $w$ in GREA features a (transient) phantom crossing\textemdash without violating the NEC\textemdash and generically predicts a slope of $w_a\simeq-0.3$, in line with observations 
    {\em Right}: $\dot{a}\equiv aH(a)$ as a function of redshift. The dashed black lines in both figures correspond to $\Lambda$CDM.}
    \label{fig:Hz1pz}
\end{figure}

To obtain the expansion history of the Universe in GREA, we numerically solve for the (rescaled) conformal time $\tau\equiv H_0\eta=H_0\int\frac{dt}{a(t)}$, using
\begin{equation}\label{eq:tau}
    \tau'\equiv\frac{d\tau}{da}=\left[a^2 \sqrt{\Omo a^{-3}(1+\frac{a_{\rm eq}}{a})+\frac{4\pi}{3a^2}\frac{\sinh{(2\tau)}}{(-k)^{3/2}V_c}}   ~\right]^{-1}~,
\end{equation}
where $a_{\rm eq}\equiv\Omega_{\rm r,0}/\Omo$ is the scale factor at matter-radiation equality, and where $\Omega_{r,0}=\Omega_{\gamma,0}+\Omega_{\nu,0}$ accounts for photons and (massless) neutrinos and $\Omo=\Omega_{\rm b,0}+\Omega_{\rm cdm,0}$ for baryons and cold dark matter, respectively. In this framework, it is possible to show (see \cite{Garcia-Bellido:2024qau}) that the comoving volume, $V_c$, is related to the spatial curvature, $k$, and the conformal time $\eta_0=\eta(z=0)$ by 
\begin{equation}
    (-k)^{3/2}V_c=\pi\left[\sinh{(2\sqrt{-k}\,\eta_0)}-2\sqrt{-k}\,\eta_0\right]~.
\end{equation}
It is the second term in \cref{eq:tau} that plays the role of the dark energy, and drives the late-time accelerated expansion of the Universe.
The phenomenology of GREA can be characterized with a single $\mathcal{O}(1)$ parameter, $\alpha$, defined as 
\cite{Garcia-Bellido:2024qau}
\begin{equation}\label{eq:alpha}
    \alpha~ \mathcal{D}_H(z=0) = \sqrt{-k}\,\eta_0~,
\end{equation}
where $\mathcal{D}_H(a)\equiv H(a)\,d_H(a)=Ha\,\eta(a)$ is the \textit{dimensionless} horizon distance, computed from \cref{eq:tau}\footnote{In \wcdm, this is given by $\mathcal{D}_H(a)=\frac{2}{\sqrt{\Omega_m}}a^{3/2}\cdot {_2F_1}\left[\frac12,\frac{-1}{6w},1-\frac{1}{6w},\frac{\Omega_m-1}{\Omega_m}a^{-3w}\right]$.}. Its value today is $\mathcal{D}_H(z=0) \neq \tau_0$.
As is clear from \cref{eq:alpha}, $\alpha$ determines the size of the curvature scale, $\sqrt{-k}\eta_0$, relative to the causal horizon today.
We integrate \cref{eq:tau} with initial conditions deep in the radiation era $\tau(a_{\rm ini}=10^{-11})=a_{\rm ini}/\sqrt{\Omega_\gamma+\Omega_\nu}$, giving $\tau(a)$. The effect of varying $\alpha$ on the \textit{effective} equation of state component, $w\equiv P/\rho$, and expansion history is shown in \cref{fig:Hz1pz}, where the corresponding values for $\sqrt{-k}\,\eta_0$ are reported in the color bar, and where the black dashed line depicts the \lcdm\ predictions.

\section{Datasets}\label{sec:DataMethods}

To derive cosmological constraints on GREA, we use the most recent Baryon Acoustic Oscillations (BAO) observations from the Dark Energy Spectroscopic Instrument (DESI) \cite{DESI:2024aqx} along with different supernovae Ia (SNe Ia) compilations and cosmic microwave background (CMB) data. Here we briefly explain the data and methods we use in this analysis. 

Our main observational data set is the baseline Baryon Acoustic Oscillations (BAO) data (DR2) from the Dark Energy Spectroscopic Instrument (DESI)
\citep{DESI2024.VI.KP7A,DESI:2024aqx,DESI:2024kob}. We combine the DESI DR2 BAO data with complementary data sets: Supernovae Ia, Cosmic Microwave Background and Big-Bang Nucleosynthesis, as described below.

\paragraph{Baryon Acoustic Oscillations (BAO)}
This dataset, abbreviated as ``DESI BAO'', spans seven redshift bins from $z=0.3$ to $z=2.33$ (where the last bin extends out to $z\approx3.5$) \cite{DESI2024.III.KP4}. DESI is an optical spectroscopic Stage IV survey carrying out a 3-D map over 14,000 square degrees and within the redshift range  $0.1<z<4.2$ \citep{Snowmass2013.Levi,DESI2016b.Instr,DESI2022.KP1.Instr,FocalPlane.Silber.2023,Corrector.Miller.2023,Spectro.Pipeline.Guy.2023,SurveyOps.Schlafly.2023,FiberSystem.Poppett.2024}. DESI was designed to improve cosmological constraints on both the expansion history and the growth rate of large scale structure, through measurements of the clustering of galaxies (BGS, LRGs and ELGs), quasars or QSOs, and the Lyman-$\alpha$ forest \citep{DESI2016a.Science,DESI2023a.KP1.SV,DESI2023b.KP1.EDR,DESI2024.IV.KP6}. In this work, we use the BAO measurements from the first 3 years of DESI observations \citep{DESI.DR2.BAO.cosmo}. The galaxy and quasar BAO measurements used here are in configuration space \citep{DESI2024.III.KP4}.

Transverse to the line of sight, DESI constrains the ratio between the comoving angular diameter distance, $D_M$, at the effective redshift of the galaxy sample, $z_{\rm eff}$, and the sound horizon at drag time, $r_{d}\equiv r_s(z_{\rm drag})$, \citep{DESI2024.III.KP4}:
    \begin{equation}\label{eq:BAO_perp}
    \frac{D_M(z_{\rm eff})}{r_{\rm d}}\equiv\frac{c}{r_{\rm d}}\int_0^{z_{\rm eff}}\frac{dz'}{H(z')}=\frac{c}{H_0r_{\rm d}}\int_0^{z_{\rm eff}}\frac{dz'}{E(z')}\ ,
    \end{equation}
where $c$ is the speed of light and $E(z)=H(z)/H_0$ is the normalized Hubble rate. The radial measurements of the BAO constrain the ratio between the Hubble distance $D_H\equiv c/H(z)$, at the effective redshift of the galaxy sample, $z_{\rm eff}$, and the sound horizon at drag epoch, $r_{d}$:
    \begin{equation}\label{eq:BAO_par}
    \frac{D_H(z_{\rm eff})}{r_{\rm d}}\equiv\frac{c}{H(z_{\rm eff})r_{\rm d}}=\frac{c}{H_0r_{\rm d}}\frac{1}{E(z_{\rm eff})}\ ,
    \end{equation} 
In the equations above, \cref{eq:BAO_par,eq:BAO_perp}, we explicitly factored out the absolute scaling set by the (degenerate) combination $H_0r_{\rm d}$. For the lowest redshift bin, corresponding to the BGS sample, we use the volume-averaged quantity:
    \begin{equation}
    \frac{D_V(z_{\rm eff})}{r_{\rm d}} \equiv [z_{\rm eff}D_M^2(z_{\rm eff})D_H(z_{\rm eff})]^{1/3}/r_{\rm d}~.
    \end{equation}

One crucial quantity for the analysis is the sound horizon, $r_s$, defined as 
\begin{equation}\label{eq:rs}
    r_s(z)=\int_{z}^{\infty}\frac{c_s(z')}{H(z')}dz'~,
\end{equation}
where $c_s(z)$ is the sound speed of the photon-baryon plasma, given by 
\begin{equation}\label{eq:cs2}
c_s^2(z)=\frac{c^2}{3(1+\frac34 R(z))}\;,\;\;\;\; R(z)\equiv \frac{\rho_b(z)}{\rho_\gamma (z)}~,
\end{equation}
and $c$ denotes the speed of light in vacuum. The sound horizon in \cref{eq:rs} is evaluated at the baryon drag epoch $\zdrag\simeq1066$, and at recombination $z_*\simeq1089$.
The exact values for \zdrag\ and $z_*$ are obtained using the approximations presented in \cite{Aizpuru:2021vhd}. These approximations provide subpercent accuracy for a wide range of parameter values, and they are further justified by the fact that at high-redshifts, GREA has a \lcdm-like expansion history. This is shown in the right panel of \cref{fig:Hz1pz}.

\paragraph{Supernovae Ia (SNe~Ia)} \label{sec:sn_data}
We use supernova (SN) data from three sets: ``\panp'', ``Union3'', and ``DES-SN5YR''. ``\panp'' is a compilation of 1550 supernovae spanning the redshift range $0.01 <z<2.26$ \cite{Brout:2022vxf}. ``Union3'' contains 2087 SNe~Ia processed through the Unity 1.5 pipeline based on Bayesian Hierarchical Modelling \cite{Rubin:2023ovl}. ``DES-SN5YR'' is a compilation of 194 low-redshift SNe~Ia ($0.025<z<0.1$) and 1635 photometrically classified SNe~Ia covering the range $0.1<z<1.3$ \cite{DES:2024tys}. These constrain the distance modulus
    \begin{equation}\label{eq:dist_mod}
        \mu(z)=m_B(z)-M_B=5\log_{10}(D_L(z)/\rm Mpc)+25~,
    \end{equation}
    where $D_L(z)=(1+z)D_M(z)$ is the luminosity distance, encoding the cosmological model-dependence. $m_B(z)$ and $M_B$ are the apparent and absolute magnitude, respectively. In this work, we treat $M_B$ as a nuisance parameter.

\paragraph{SH0ES measurement of $H_0$} 
The \panp~  sample includes a set of very low-redshift Type Ia supernovae with Cepheid-based distance calibrations \cite{Brout:2022vxf,Riess:2021jrx}. These serve as the local anchor in the distance ladder and effectively constrain the Hubble constant, $H_0$, or equivalently, the absolute magnitude of Type Ia supernovae, $M_B$. In practice, we use the \texttt{pantheonplusshoes}\footnote{\url{https://cobaya.readthedocs.io/en/latest/likelihood_sn.html}} likelihood, as implemented in \cobaya\footnote{\url{https://github.com/CobayaSampler/cobaya}}, and refer to this as \panp~\& \shoes.

\paragraph{Cosmic Microwave Background (CMB)} 

Following \cite{DESI:2025zgx,DESI:2025fii}, we use the compressed CMB information in terms of $(\theta_s,\omega_{\rm b},\omega_{\rm cb})$. These have been found to yield nearly identical constraints on late-time dark energy as the full CMB likelihood. 
The mean values and covariance matrix have been extracted from the publicly  available\footnote{The chains can be found at \url{https://act.princeton.edu/act-dr6-data-products}} P-ACT \lcdm\ chains \cite{ACT:2025fju,ACT:2025tim}, and include both Planck PR4 \cite{Carron:2022eyg} and ACT DR6 lensing \cite{ACT:2023kun,ACT:2023dou}).


\paragraph{Big-Bang Nucleosynthesis (BBN)} 
When we don't include the CMB data, we include a prior on $\omega_b\equiv\Omega_bh^2$ from big-bang nucleosynthesis (BBN) \cite{Schoneberg:2024ifp} to break the $H_0$-$r_{\rm d}$ degeneracy and ``calibrate'' the distances from BAO and SNe.

\section{Methodology}\label{sec:methods}
\begin{table}[t]
    \caption{
    Parameters and priors used in the analysis. All of the priors are uniform in the ranges specified below, except for the runs including $+\rm BBN$, where a Gaussian prior on $\omega_b$ is used instead \cite{Schoneberg:2024ifp}. 
    }\label{tab:priors} 
    \centering
    \begin{tabular}{|c|c|}
    \hline
     parameter & prior\\  
    \hline 
    $\omega_{\rm cdm}\equiv\Omega_{\rm cdm,0}h^2$ &  $\mathcal{U}[0.05, 0.5]$\\
    $\omega_{\rm b}\equiv\Omega_{\rm b,0}h^2$ &  $\mathcal{U}[0.005, 0.1]$ \\
    $H_0 \; [\kmsMpc]$ &  $\mathcal{U}[20, 100]$  \\
    $\sqrt{-k}\eta_0$ &  $\mathcal{U}[1, 5]$  \\
    $\Omega_{\rm k,0}$ &  $\mathcal{U}[-0.3, 0.3]$  \\
    \hline
    \end{tabular}
\end{table}

To confront the predictions of GREA to the data, we numerically implement \cref{eq:tau} into a Python package, \texttt{greapy}\footnote{\url{https://github.com/rcalderonb6/greapy}}. We then sample the posterior distribution of GREA using Markov Chain Monte Carlo (MCMC) methods. In particular, we use the Metropolis-Hastings algorithm \cite{Lewis:2002ah,Lewis:2013hha}, as implemented in the publicly available sampler \cobaya~\cite{Torrado:2020dgo}. The parameter space associated with GREA is $\Theta=\{H_0,\omega_{\rm cdm},\omega_b,\sqrt{-k}\eta_0\}$ and unless specified otherwise, we assume flat uninformative priors on all parameters, listed in \cref{tab:priors}. Throughout this work, we assume massless neutrinos, with $N_{\rm eff}=3.044$ and a fixed spectral index, $n_s = 0.965$. We do not expect the inclusion of massive neutrinos or a different spectral tilt to significantly alter our conclusions, as none of the observables used in the analysis are sensitive to such parameters. 
We assess the convergence of our MCMC chains using the Gelman-Rubin ($R-1$) criterion \cite{1992StaSc...7..457G} and require all of our chains to satisfy $R-1<0.01$. We analyze our chains and produce our plots using the python package \gd~\cite{Lewis:2019xzd}. Finally, for the \panp, DES-SN5YR, and Union3 likelihoods, the marginalization over the absolute magnitude $M_B$ is done analytically.
As in previous DESI analyses, we adopt the $\alpha_\perp$ and $\alpha_\parallel$ parametrization to characterize the angular and radial BAO scales relative to a fiducial $\Lambda$CDM model. While a fully consistent treatment would involve recomputing the anisotropic two-point correlation function within the GREA framework, this is beyond the scope of the present work. Nonetheless, we note that the standard BAO compression scheme has been shown to be robust for a wide class of cosmological models, including those beyond $\Lambda$CDM \cite{Pan:2023zgb}. In particular, GREA belongs to the class of homogeneous and isotropic FLRW models, and its linear perturbation theory yields a growth history (\cref{fig:growth}) that encompasses that of $\Lambda$CDM~\cite{Garcia-Bellido:2024qau}. The primary difference lies in the dynamical evolution of the effective dark energy component.

To compute the Bayesian Evidence $\mathcal{Z}$ required for model selection, we use the nested-sampling algorithm implemented in \textsc{polychord} \cite{Handley:2015fda,Handley_2015} for specific data combinations. In essence, Bayesian Evidence enables the direct comparison of competing models via their ratio, known as the Bayes factor, $K$, which quantifies the relative support for one model over another by simultaneously considering the goodness of fit and penalizing for model complexity. However, it is important to note that model selection based on the Bayes factor has limitations, and conclusions drawn from such analyses must be interpreted with caution \cite{Nesseris_2013,Koo_2022,10.1093/mnras/stac1851}. The interpretation of the Bayes factor can be sensitive to the choice of priors, not only parameter ranges but also theoretical priors, and its effectiveness may diminish when comparing models that differ significantly in their underlying theoretical assumptions or parameter space.

\section{Results and Discussions}\label{sec:results}

\subsection{Late-time probes}
\begin{figure}[t]
    \centering
    \includegraphics[width=\textwidth]{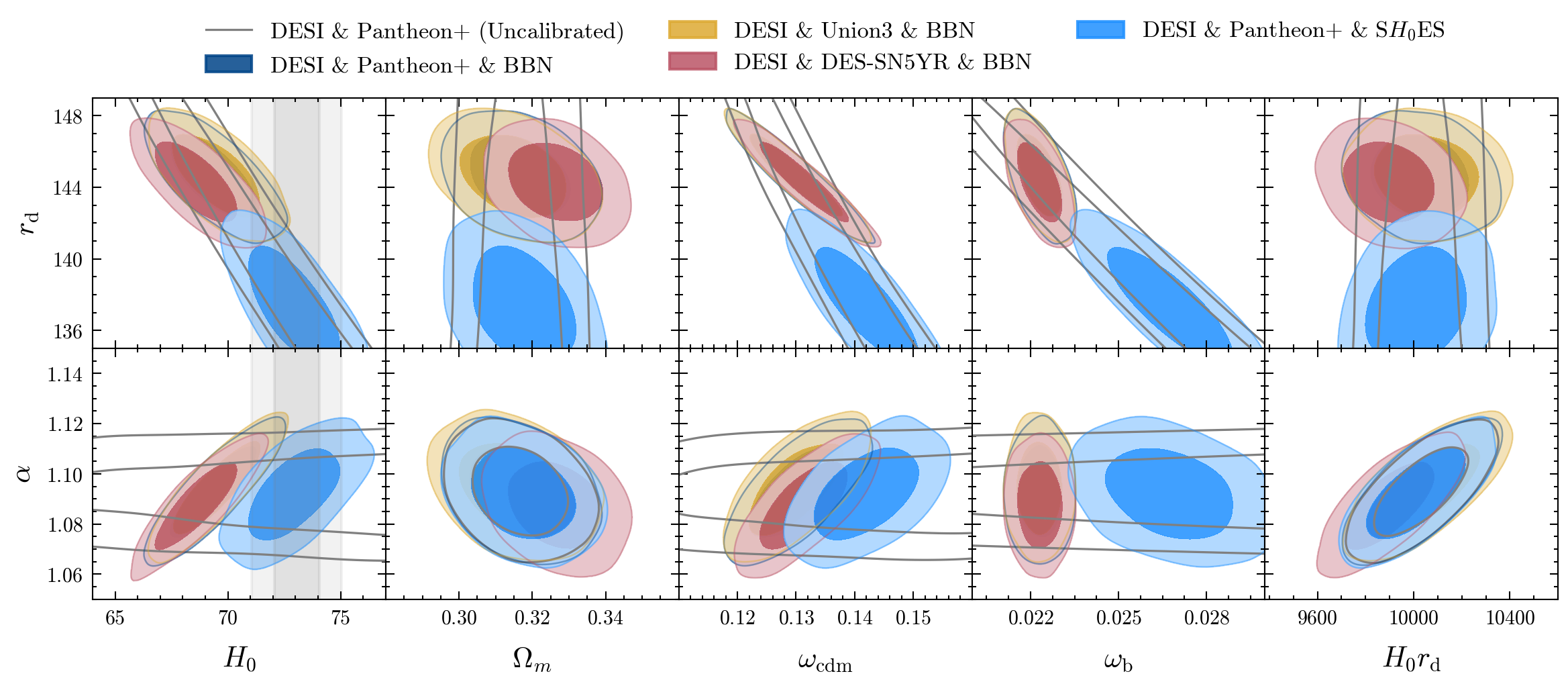}
    \caption{Marginalized constraints on parameters of GREA from DESI BAO DR2 and the various SNe Ia compilations: ``\panp'', ``Union3'', and ``DES-SN5YR''  (\cref{sec:sn_data}). We show the impact of an early-universe calibration of distances through the inclusion of a BBN-prior on $\omega_b\equiv\Omega_bh^2$ \cite{Schoneberg:2024ifp}, or at late-times by including \shoes \cite{Riess:2021jrx}. The light gray unfilled contours show the uncalibrated results when using the DESI BAO and \panp~datasets, which constrain only the product $H_0r_{\rm d}$ and $\Omo$. For reference, we show here the $H_0$ measurements from \shoes \cite{Riess:2021jrx} with the light grey vertical bands. The inferred parameters are consistent for the \panp, Union3, and DES-SN5YR SN datasets.
    }
    \label{fig:DESI+BBN+SNe}
\end{figure}

We begin by discussing the results obtained from late-time (low-redshift) probes.
\cref{fig:DESI+BBN+SNe} presents the constraints derived from combining the DESI DR2 BAO measurements with various available supernova Ia (SNe Ia) compilations (\cref{sec:sn_data}). The light gray (unfilled) contours represent the uncalibrated DESI BAO and \panp~datasets, which constrain only the fractional matter density, $\Omo$, and the absolute scaling, $H_0r_{\rm d}$. The uncalibrated distances probed by the SN+BAO combination constitute a sensitive probe of the shape of the expansion history $E(z)\equiv H(z)/H_0$, as seen from \cref{eq:dist_mod,eq:BAO_perp,eq:BAO_par}. Thus, the \textit{uncalibrated} SN+BAO data alone can constrain the fractional matter density, \Omo, and  $\alpha$, but not the individual physical densities $\omega_{\rm b}$ and $\omega_{\rm c}$.

The late-time probes (SN \& BAO) allow for values of $H_0$ compatible with \shoes \cite{Riess:2021jrx} (shown as a vertical shaded band in \cref{fig:DESI+BBN+SNe}) while implying a reduced size of the sound horizon, relative to \lcdm, and therefore an increase in $\omega_{\rm b}$ (in tension with BBN).
Such a value of $H_0$, in combination with a large $\Omo$, inevitably leads to an increase of $\omega_m=\omega_{\rm cdm}+\omega_{\rm b}=\Omo h^2$ (\cref{fig:DESI+BBN+SNe}). This is common to many models attempting to address the Hubble tension, as recently emphasized in \cite{Poulin:2024ken,Pedrotti:2024kpn}. Such an increase in the physical matter density has important implications for the acoustic peaks in the cosmic microwave background and the clustering of matter at late times.

To break the $H_0$-$r_{\rm d}$ degeneracy, we also incorporate a big-bang nucleosynthesis (BBN) prior on $ \omega_b=(2.231\pm0.05)\times 10^{-2}$ \cite{Schoneberg:2024ifp}, effectively calibrating distances in the early universe by fixing the size of the standard ruler, $r_{\rm d}$. The constraints on $\alpha=1.09\pm 0.012$ are stable across different SN compilations, while the constraints on $\Omo$ marginally depend on the choice of SN, with DES-SN5YR driving the preference for higher matter densities with respect to \panp~ and Union3, in agreement with previous findings \cite{DES:2024tys,DESI2024.VI.KP7A,DESI:2024aqx,DESI:2024kob}. For conciseness, in what follows we will focus on and only show results using the \panp~  compilation.

\subsection{The role of the CMB}
\begin{figure}
    \centering
    \includegraphics[width=\linewidth]{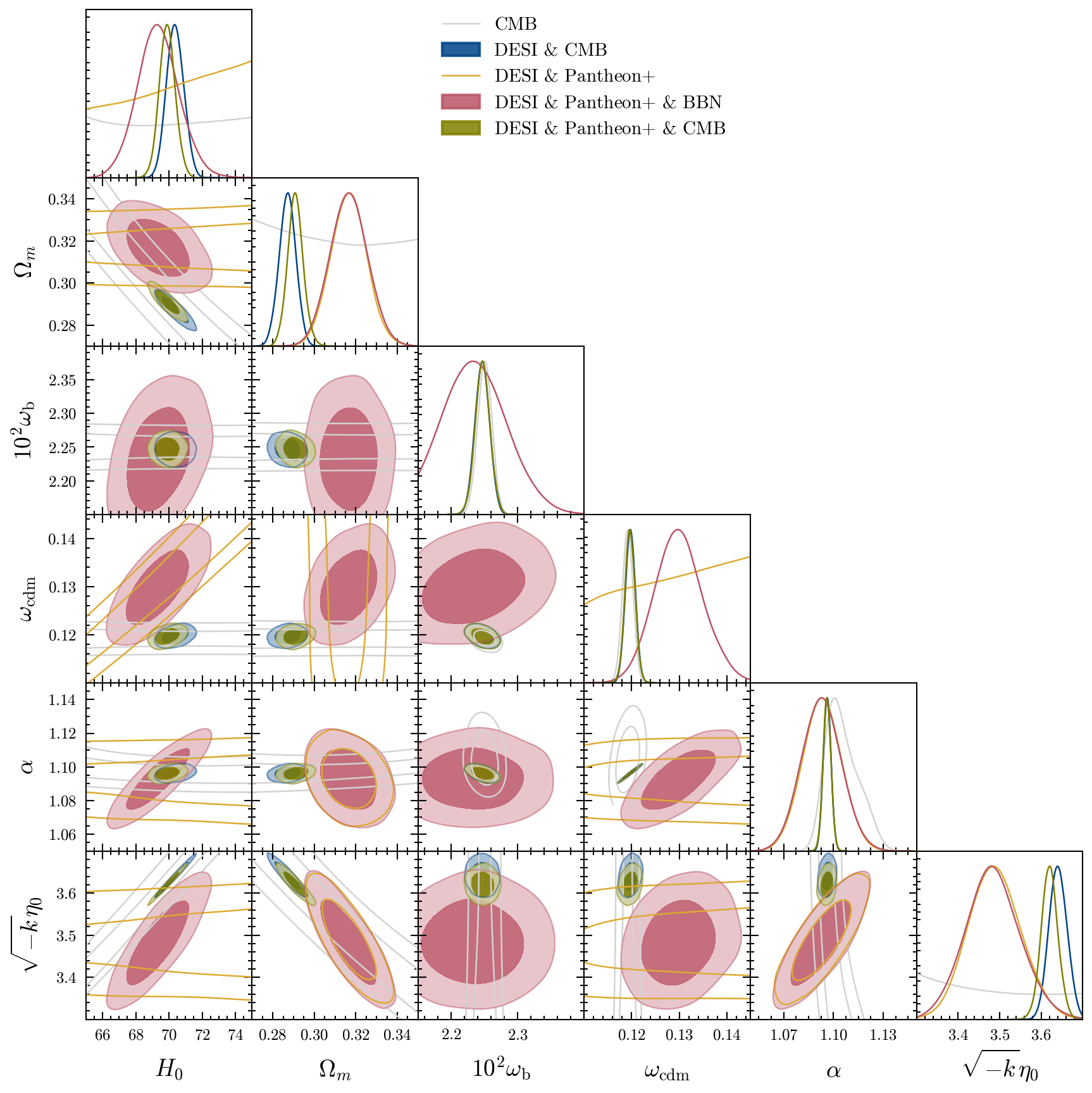}
    \caption{The full posterior distribution for GREA's cosmological parameters (\cref{sec:GREA}), inferred from fitting the model to different datasets, including DESI DR2~\citep{DESI:2025zgx}, as indicated in the legend. The light gray unfilled contours show the uncalibrated results when using the CMB data from P-ACT (\cref{sec:DataMethods}), which is sensitive to the product $\Omo h^2$. Note that the \panp~data prefer higher matter densities than the DESI \& CMB datasets, leading to some tension in some of the projected parameter space. These inconsistencies are also reflected in the degraded fit to the data.}\label{fig:DESI+CMB+PantheonPlus}
\end{figure}

\begin{figure}
    \centering
    \includegraphics[width=\textwidth]{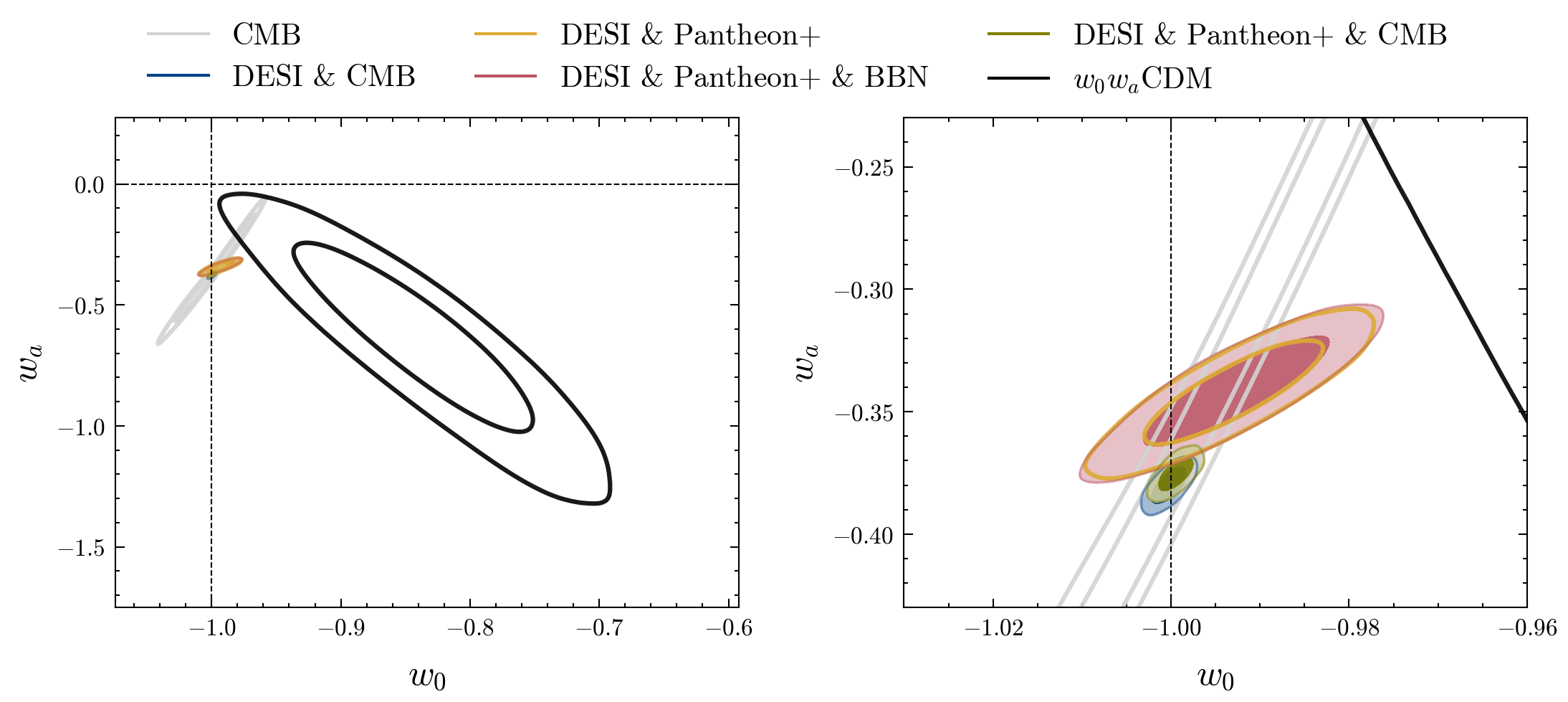}
\caption{\textit{Left}: Derived constraints on the equation of state parameter $w(a=1)$ and its derivative $w_a=-\frac{dw}{da}(a=1)$ today for GREA. Note that the parametrization $w(a)=w_0+w_a(1-a)$ does not capture the full behavior of the equation of state in GREA (\cref{fig:Hz1pz}). These two parameters only capture the very low-$z$ behavior, and we report them here for comparison with the standard $w_0$-$w_a$ analysis. A negative slope $w_a\sim-0.3$ is quite a generic prediction from GREA, while its value today $w_0$, depends on the value of $\alpha$, as illustrated in \cref{fig:Hz1pz}. \textit{Right}: Zoomed in version around the region favoured by GREA. 
}\label{fig:w0wa_posterior}
\end{figure}

In \cref{fig:DESI+CMB+PantheonPlus}, we report the constraints including CMB information. The CMB constrains the late-time expansion history primarily through an exquisite ($\sim0.04\%$) measurement of the acoustic scale $\theta_s\equiv r_s(z_*)/D_M(z_*)$. Recall that the relative height of odd/even acoustic peaks in the CMB tightly constrains the physical baryon density, $\omega_{\rm b}=\Omega_{\rm b,0}h^2$  \cite{Chu:2004qx,Motloch:2020lhu}. Thus, the angular size of the sound horizon, together with the physical energy densities $\omega_{\rm b}$ and $\omega_{\rm cdm}$, carries most of the geometrical information contained in the CMB.  Within \lcdm, the CMB distance priors provide constraints on the fractional matter density, \Omo, and Hubble constant, $H_0$. In GREA, however, because of the degeneracies introduced by the parameter $\alpha$, we can no longer estimate $\Omo$ and $H_0$ purely from such measurements (as shown by the gray contours in \cref{fig:DESI+CMB+PantheonPlus}). This is reminiscent of the intrinsic geometrical degeneracy between matter and dark energy (\Omo\ and $w(z)$) \citep[e.g.][]{Wasserman:2002gb,PhysRevD.80.123001,Shafieloo:2011zv,von_Marttens_2020,Calderon:2022cfj}. 

Let us now discuss the results of the DESI BAO \& CMB combination. While the CMB is sensitive to the product $\Omo h^2$, DESI BAO measurements probe the fractional matter density $\Omo$ directly. Thus, including the DESI data breaks these degeneracies by providing constraints on $\Omo$ (see \autoref{fig:w0wa_posterior}). In turn, this allows for an estimation of $H_0$, and more importantly, of $\alpha$, which characterizes the phenomenology of GREA. Including SNe Ia distance measurements, which are also highly sensitive to the fractional matter density, yield constraints of $\Omo = 0.317 \pm 0.009$ ($\Omo = 0.292 \pm 0.004$) for the DESI BAO \& \panp~\&  BBN (DESI BAO \& \panp~\&  CMB) combination. This increase in $\Omo$ ultimately leads to a lower estimate of $\alpha$, shifting the central value of $H_0$ towards smaller values, around $H_0 \sim 70$ for any analysis including both \panp~ and DESI BAO data. Notably, swapping \panp~ with the DES-SN5YR compilation pushes $H_0$ to even lower values, due to the preference for a higher $\Omo$ in comparison to \panp.

These differences in $\Omo$ between analyses with and without \panp~  are reflected in the degraded fit quality to the combined data, as shown in \cref{tab:best-fit}. This fit degradation, driven by subtle differences in preferred $\Omo$, is a key reason why many late-time modifications to the expansion history struggle to resolve the ``cosmic calibration'' problem \cite{Schoneberg:2021qvd,PhysRevLett.131.111002,Poulin:2024ken}, especially when combining the Hubble diagrams probed by SNe Ia and BAO.

\begin{table}[t]
    \centering
    \caption{The goodness of fit for different models and dataset combinations, $\Delta\chi^2$, obtained with the minimizer \textsc{iMinuit} \cite{iminuit}. The second column shows $\dchisq\equiv \chisq_{\rm GREA}-\chisq_{\Lambda\rm CDM}$. Note that CMB here refers to Planck PR4. 
    }  
$\,$\\ 

\label{tab:best-fit}
    \resizebox{0.95\textwidth}{!}{\begin{tabular}{|c|c|c|c|c|c|} \hline  \hline 
         Data & $\Delta \chi^2$&  $\chi^2_{\rm GREA}$ &  $\chi^2_{\Lambda\rm CDM}$ &  $\chi^2_{\Lambda\rm CDM+\Omega_k}$ & $\chi^2_{w_0w_a}$ \\  \hline  
         DESI \& BBN & $3.6$ & $13.9$ & $10.3$ & $9.9$ & $5.6$  \\
         DESI \& CMB & $3.0$ & $17.0$ & $13.9$ & $10.5$& $7.3$  \\
         DESI \& \panp~\&  BBN & $5.9$ & $1422.1$ & $1416.2$ & $1414.4$ & $1411.3$ \\ 
         DESI \& \panp~\&  PR4 & $13.4$ & $1433.5$ & $1420.1$ & $ 1416.0$ & $1413.0$ \\ 
          DESI \& \panp~\& \shoes & $5.8$ & $1471.2$ & $1465.3$ & $1463.6$ & $1460.5$ \\ 
          DESI \& \panp~\&  PR4~\& \shoes & $0.4$ & $1498.2$ & $1497.8$ & $1490.1$ & $1490.4$ \\ 
         \hline \hline 
    \end{tabular}}
\end{table}
\begin{figure}[!h]
    \centering
    \includegraphics[width=0.48\linewidth]{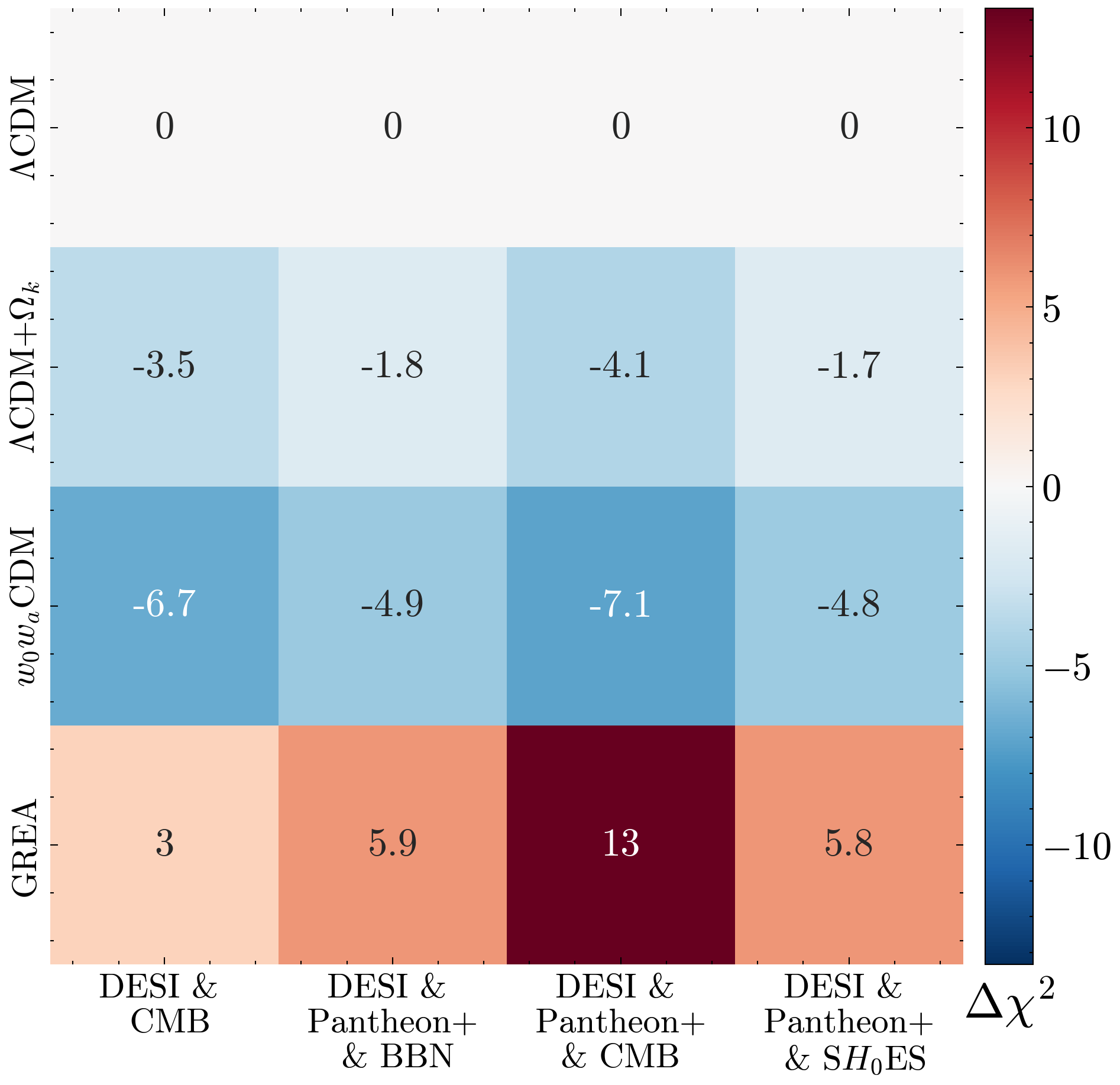}
    \includegraphics[width=0.475\linewidth]{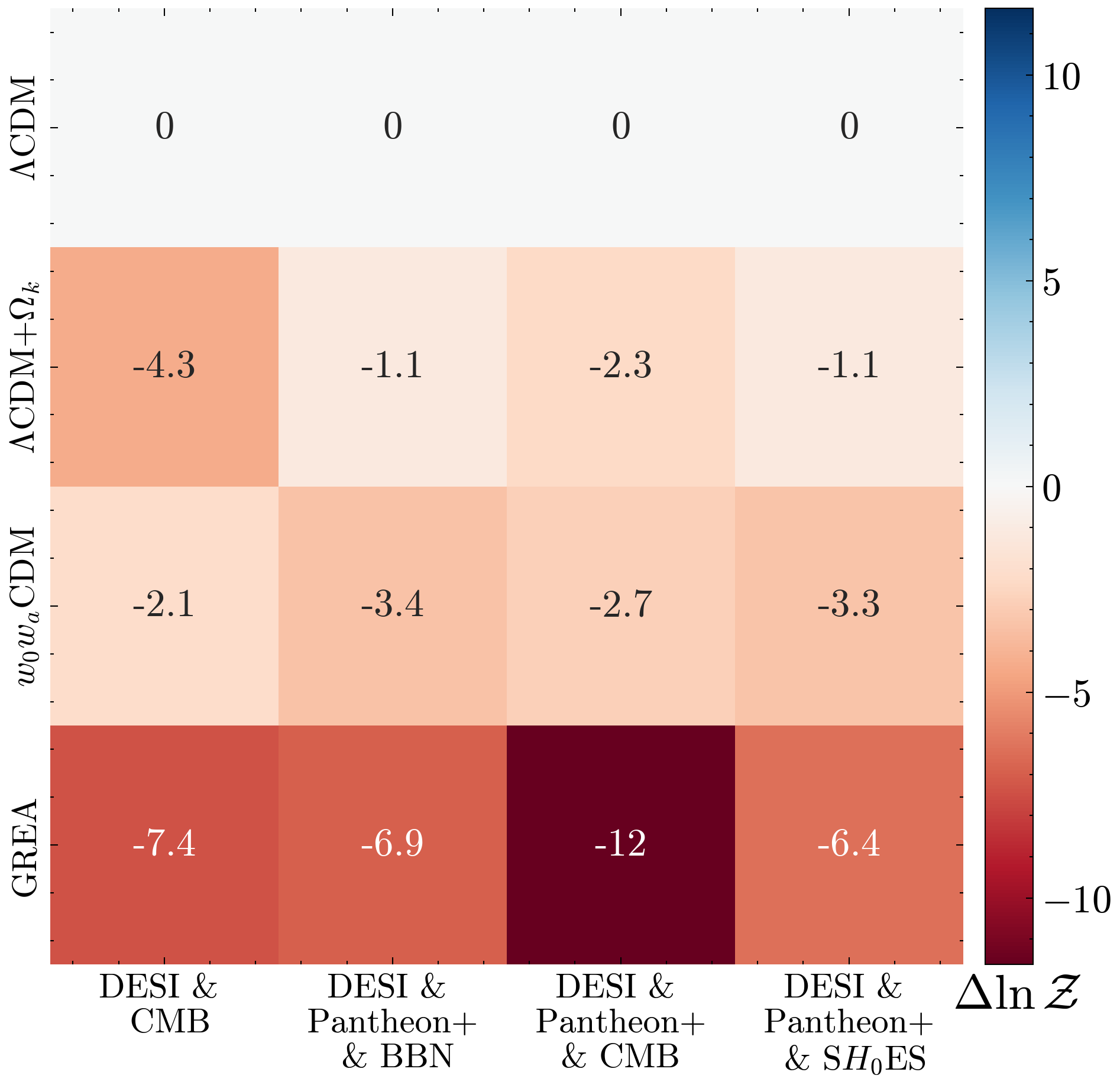}
    \caption{Visual representation of the goodness of fit compared to the \lcdm model, for the different models and dataset combinations as shown in \cref{tab:best-fit}: $\dchisq\equiv \chisq_{\rm i}-\chisq_{\Lambda\rm CDM}$ (left), and $\Delta\ln{\mathcal{Z}}=\log\mathcal{Z}^{\rm i}-\log\mathcal{Z}^{\Lambda \rm CDM}$ values (right). The Bayesian evidence, $\mathcal{Z}$, is computed using \textsc{polychord}~\cite{Handley_2015}.}\label{fig:heatmap_Fit}
\end{figure}

Finally, it is insightful to introduce an \textit{effective} dark energy density $\rho^{\rm eff}_{\rm DE}$ as
\begin{equation}
    \Omega_{\rm DE,0}\ f_{\rm DE}(z)\equiv \left(\frac{H(z)}{H(0)}\right)^2-\Omo (1+z)^3-\Omega_{\rm r,0}(1+z)^4~,
\end{equation}
where $\Omega_{\rm DE,0}=\frac{8\pi G }{3H(0)^2}\rho^{\rm eff}_{\rm DE,0}\simeq(1-\Omo)$ and $f_{\rm DE}(z)=\rho^{\rm eff}_{\rm DE}(z)/\rho^{\rm eff}_{\rm DE,0}$ determines its redshift evolution, from which we can compute its \textit{effective} equation of state, using
\begin{align}
    w(a) =-\frac13\frac{\mathrm{d}\ln{f_{\rm DE}}}{\mathrm{d}\ln{a}}-1&=-\frac13\frac{\mathrm{d}\ln}{\mathrm{d}\ln{a}}\left[\frac{\sinh(2\tau)}{a^2}\right]-1\\
    &= -\frac13\bigg[2a\,\tau'(a)\coth[2\tau(a)]+1\bigg] \,.
\end{align}

\begin{figure}[h]
    \centering
\includegraphics[width=1\textwidth]{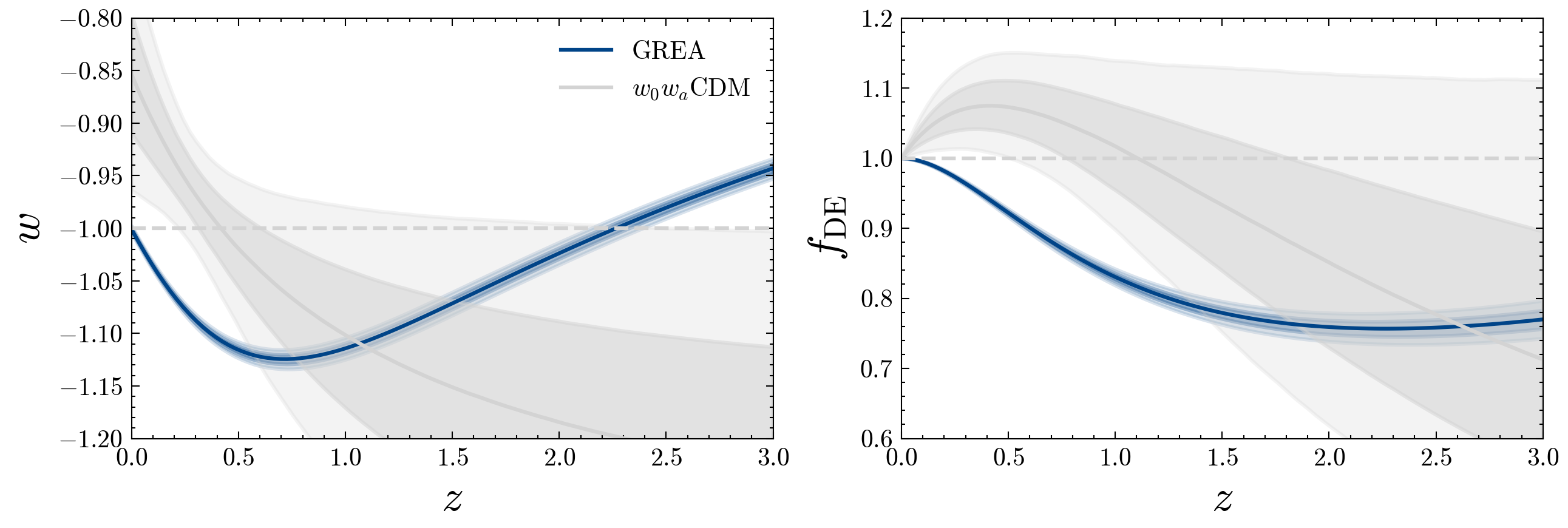}
    \caption{Constraints on the equation of state, $w$, and normalized energy density, $\fde=\rho_{\rm DE}/\rho_{\rm DE,0}$, from the DESI \& \panp~\&  CMB data combination in $w_0w_a\rm CDM$ (in gray) and GREA (in blue) The shaded bands represent the $68\%$ and $95\%$ confidence levels computed by drawing a large number of samples from the chains. 
    }
    \label{fig:ev_w_rho}
\end{figure}
The marginalized constraints on these two key quantities are depicted in \cref{fig:ev_w_rho} for the DESI \& \panp\ \& CMB data combination. These two quantities allow for a direct comparison of the best-fit GREA model with the data-driven (model-agnostic) reconstructions of the DE phenomenology, presented in \cite{DESI:2024aqx,Y3.cpe-s1.Lodha.2025,DESI:2025wyn}.
The behaviour of $w(z)$ in GREA, particularly at low redshift, does not fully match the trends suggested by non-parametric, model-independent reconstructions. This mismatch is also reflected in the fit to the data, where the phenomenological $w_0w_a\mathrm{CDM}$ model\textemdash whose functional form closely follows the reconstructed $w(z)$\textemdash consistently outperforms both $\Lambda$CDM and GREA. While this discrepancy could have been anticipated from the allowed shape of $w(z)$ in GREA (\cref{fig:Hz1pz}) even before fitting, it is still valuable to confront the model with data, as this allows us to constrain its parameters and quantitatively assess its viability.

\begin{figure}
    \centering
    \includegraphics[width=\linewidth]{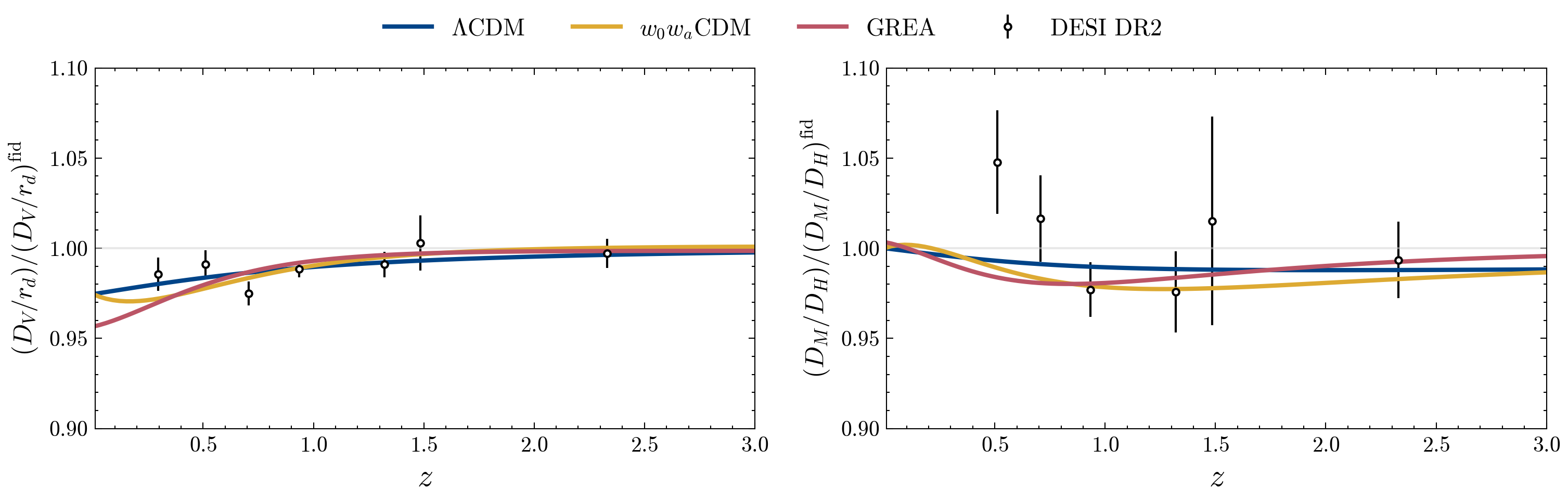}
    \caption{Evolution with redshift of the volume-averaged distance, $D_V$ (left panel); and the anisotropic distortion measured with the comoving angular diameter distance signal, $D_M$ (right panel, see details in \cref{sec:DataMethods}). Observational datasets are shown as open symbols. In each panel, the best fit model (BAO \& SNeIa \& CMB) for \lcdm is shown by blue lines, GREA by red lines, and the \wowacdm model by yellow lines. The largest differences between the GREA and \wowacdm\ bestfit models are found at $z<1$.
    }\label{fig:dm_dv}
\end{figure}

In \cref{fig:dm_dv}, we display the best-fit predictions for the different models on top of BAO observables $D_V(z)/r_{\rm d}$ and $F_{\rm AP}(z)\equiv D_M/D_H$ normalized to their fiducial values\footnote{This corresponds to the baseline \lcdm\ parameters used by the AbacusSummit suite of N-body simulations \url{https://github.com/abacusorg/AbacusSummit/blob/main/Cosmologies/abacus_cosm000/CLASS.ini}. For more details on the BAO measurements, see e.g. \cite{DESI2024.II.KP3}.}. The isotropic and anisotropic BAO distortion parameters at $0<z<1$ can help distinguish the best fit GREA and the \wowacdm\ models (see  \cref{fig:dm_dv}). 

\begin{figure}
    \centering
    \includegraphics[width=\linewidth]{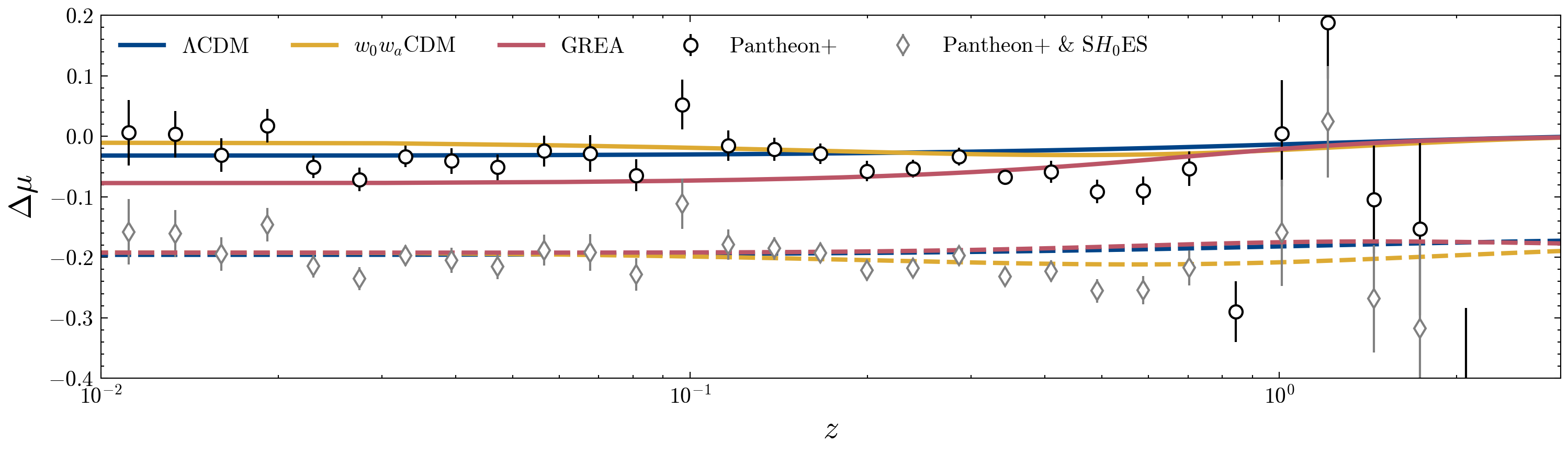}
    \caption{Evolution with redshift of the distance modulus residuals, $\Delta\mu=\mu-\mu^{\rm fid}$. The (uncalibrated) \panp~ are shown as open symbols, where as the \shoes-calibrated \panp\ measurements are shown as diamonds. The solid lines correspond to the best fit models to the full (DESI \& \panp \& CMB) data combination. The dashed lines shows the corresponding bestfit predictions to the \shoes-calibrated (DESI \& \panp) measurements.
    }\label{fig:mu}
\end{figure}

When fitting the combined (DESI \& \panp\ \& CMB) data, the high-redshift evolution of the distance modulus is the same for both the best-fit GREA and \wowacdm\ models describing current observations, except for very low redshift $z<0.5$, where the increase in $H^{\rm GREA}>H^{\Lambda\rm CDM}$ leads to $D_L^{\rm GREA}<D_L^{\Lambda\rm CDM}$ (\cref{fig:mu}). Note that the uncalibrated \panp\ data (black circled datapoints) does not constrain $H_0$, nor $M_B$, but only their product. Thus, for a given $H_0$ there is freedom to choose the corresponding $M_B$ that best fits the data. For simplicity, we normalize these measurements to the fiducial \lcdm\ value.
When calibrating distances with the cepheids-based (\shoes) measurements (dashed lines in \cref{fig:mu}), as expected, the very low-$z$ evolution is almost identical between the three models, and deviations only appear at intermediate redshift, where the $w_0w_a\rm CDM$ seems to fit the SN data better.

\subsection{Further constraining GREA}

\begin{figure}[t]
    \centering
    \includegraphics[width=\linewidth]{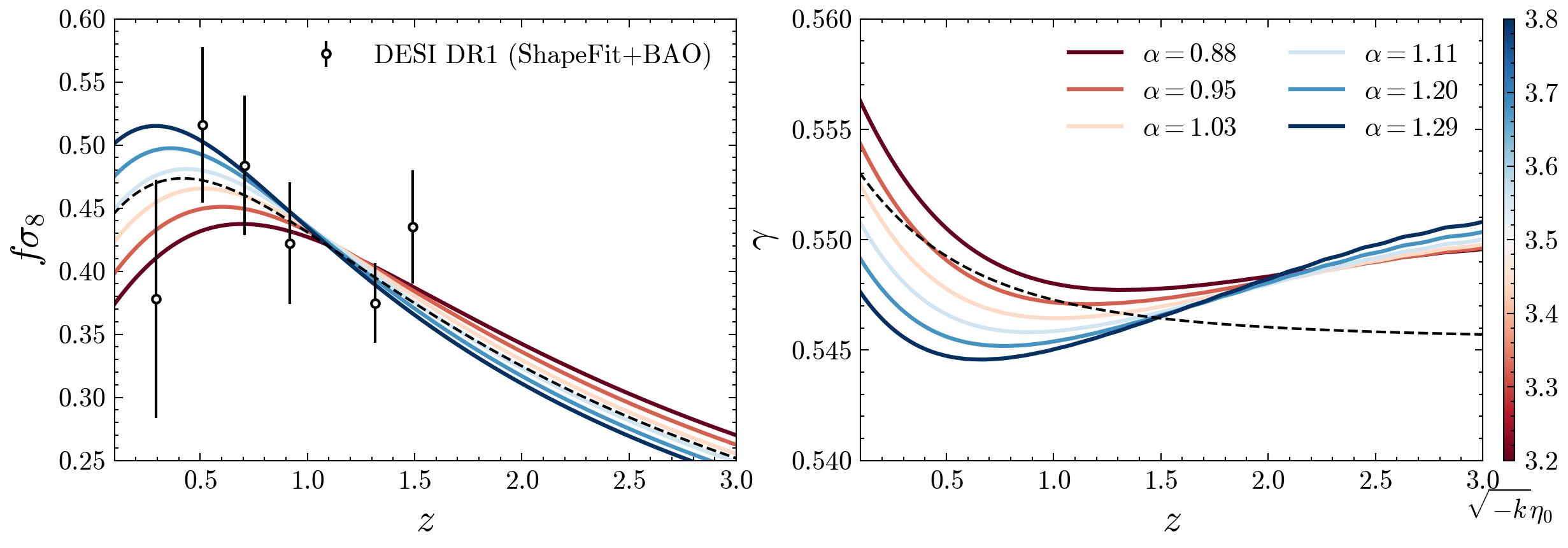}
    \caption{Growth of density perturbations as a function of redshift for various values of $\alpha$ in GREA. \textit{Left:} Evolution of the combination $f\sigma_8(z)$, probed by redshift-space distortions. \textit{Right:} Corresponding growth index $\gamma(z)\equiv \ln f(z)/\ln\Omega_m(z)$ evolution. The $f\sigma_8(z)$ measurements have been extracted from the DESI DR1 ShapeFit+BAO analysis~\cite{DESI:2024jxi}. They are for illustrative purposes only and have not been used in the present analysis. The black dashed line on both panels corresponds to \lcdm.}
    \label{fig:growth}
\end{figure}

In GREA, for scales of cosmological interest, the growth of density perturbations (in the linear regime) obeys \cite{Garcia-Bellido:2024qau}
\begin{equation}\label{eq:growth_d}
a(\tau)\delta''(\tau) + a'(\tau)\delta'(\tau) = \frac{3}{2} \Omega_{\rm m,0}\delta(\tau)\,, 
\end{equation}
where $\delta=\rho/\bar\rho-1$ is matter density contrast and a prime denotes differentiation with respect to conformal time, $\tau$. 
The modified growth of perturbations stems from the modified evolution of the scale factor, $a(\tau)\neq a^{\Lambda \rm CDM}$, encoded in the Hubble friction term above. 
It is convenient to rewrite the growth equation \cref{eq:growth_d} in terms of the growth factor $f\equiv\delta'/\delta=\frac{d\ln\delta}{d\ln a}$, giving 
\begin{equation}\label{eq:growth_f}
f'(a) + \left(f(a)+2+\frac{E'}{E}\right) f(a) = \frac 3 2 \Omega_{\rm m}(a)~,
\end{equation}
where $^{\prime}\equiv d/d\ln a$, $\Omega_{\rm m}(a)=\Omo\ a^{-3} E^{-2}$ and $E=H(z)/H_0$. We integrate \cref{eq:growth_f} with initial conditions deep in the matter dominated epoch, with $f(a_{\rm ini})=1$\footnote{In an Einstein-DeSitter (EdS), or matter dominated universe, it is easy to show that $\delta(a)\propto a\Leftrightarrow f=1$.}. From an observational viewpoint, the relevant quantity is the combination \cite{Song:2008qt}
\begin{equation}
f\sigma_8(a)\equiv f(a)\cdot\sigma_8(a)=\frac{\sigma_{8,0}}{\delta_0} f\delta = \frac{\sigma_{8,0}}{\delta_0}\delta',
\end{equation}
where $\sigma_{8,0}$ is the amplitude of matter fluctuations, smoothed over scales of radius $R=8h^{-1}\rm Mpc$, and $\delta_0=\delta(z=0)$. On the left panel of \cref{fig:growth}, we show the evolution of the growth rate $f\sigma_8(a)$ at late times for various GREA scenarios, along with the DESI DR1 measurements extracted from a ShapeFit+BAO analysis (Tab. 11 in  \cite{DESI:2024jxi}).
The present data constrain the background expansion history in GREA to be close to that of \lcdm, with preferred values of $\alpha\sim1.09$, seen in \cref{fig:DESI+CMB+PantheonPlus,tab:parameter_table2}. Therefore, the corresponding $f\sigma_8$ closely follows that of \lcdm, as the growth history is fully determined by the expansion history for DE models inside GR. On the right panel, we display the evolution of the growth index $\gamma(z)\equiv\ln{f}(a)/\ln{\Omega_m(a)}$~\cite{1976ApJ...205..318P,1991MNRAS.251..128L,Wang:1998gt,Linder:2005in,Polarski:2007rr}. Interestingly, while the present value of $\gamma\simeq 0.55$ is close to that of \lcdm, at higher redshifts, $\gamma(z)$ has a very different behaviour, satisfying in particular $d\gamma/dz>0$. This is in sharp contrast with the \lcdm\ case, where $\gamma(z)$ is a monotonically decreasing function of redshift \cite{Polarski:2016ieb,Calderon:2019jem,Calderon:2019vog}. 
Forthcoming LSS measurements by DESI, in combination with Euclid and LSST, are expected to constrain $\gamma$ and its derivative with high precision \cite{Amendola:2016saw}.

\section{Conclusions}\label{sec:conclusions}
General Relativistic Entropic Acceleration (GREA) arises as a consequence of out of equilibrium phenomena associated with the growth of entropy of horizons. It is a consequence of quantum gravity holography, which gives a correspondence between the dynamics in a volume of space and the degrees of freedom in the boundary of that space. The GREA theory offers a natural explanation for the universe's accelerated expansion without the need for a cosmological constant~\citep{Garcia-Bellido:2021idr,Arjona:2021uxs,Garcia-Bellido:2024tip,Garcia-Bellido:2024qau}. The accelerated expansion in GREA is associated with the growth of entropy from cosmic and black hole horizons, making it a promising alternative to the cosmological constant, $\Lambda$. In addition, this theory introduces a single $\mathcal{O}(1)$ parameter $\alpha$ {that encapsulates the effect of GREA as an evolving dark energy scenario, both at the background and perturbation level}. GREA makes precise predictions for various cosmological observables that can be tested with increasingly accurate observations \cite{Garcia-Bellido:2024qau}. For our Universe, the causal horizon growth associated with the matter/energy content induces a growth in entropy of the horizon while driving a dynamical entropic acceleration, which is very different from that predicted by \lcdm. Even in the absence of $\Lambda$, the entropic acceleration is enough to explain the dimming of distant SNe and the recent BAO data from DESI DR2 (\cref{fig:dm_dv}). Moreover, GREA makes a specific prediction: that the size of our causal horizon today should be the same as the curvature scale ($\alpha\sim 1$).

In this work, we have used state-of-the-art BAO measurements by the DESI collaboration \citep{DESI2022.KP1.Instr,DESI2024.II.KP3,DESI2024.III.KP4,DESI2024.IV.KP6,DESI2024.VI.KP7A},
together with other external datasets (\cref{sec:DataMethods}), to constrain the free parameters of GREA (\cref{sec:methods}). 
GREA achieves a fit to observations comparable to the phenomenological \lcdm\ model\footnote{We note that $\chi^2$ differences are reported relative to \lcdm. The goodness of fit of \lcdm\ has been assessed in previous works \cite{DESI:2024jxi,DESI.DR2.BAO.cosmo}.} (\cref{tab:best-fit} and \cref{fig:heatmap_Fit}). This is particularly encouraging as \lcdm\ and GREA are not nested models. Despite the Bayesian evidence favouring the phenomenological \lcdm\ model, the alternative GREA scenario remains compelling. 
We emphasize that usual model comparison metrics, such as $\Delta\chi^2$ and the Bayes ratio, should be interpreted with care for non-nested models, such as GREA and \lcdm.
We have reported the Bayes ratios assuming equal priors for the underlying models. Perhaps a more appropriate choice is to weight the alternative hypothesis according to their theoretical basis. This could potentially 
penalize \lcdm\ and \wowacdm\ with respect to GREA and change the odds in its favour. In the Bayesian framework, the preference for one model ($M_1$) over another ($M_0$) is usually assessed by computing the Bayes ratio, $K$,  given by 
\begin{equation}
    K=\frac{\mathcal{Z}_1}{\mathcal{Z}_0}=\frac{\mathcal{P}(D|M_1)}{\mathcal{P}(D|M_0)}=\frac{\mathcal{P}(M_1|D)}{\mathcal{P}(M_0|D)}\frac{\Pi(M_0)}{\Pi(M_1)}~,
\end{equation}
where $\Pi(M_i)$ and $\mathcal{Z}_i=\mathcal{P}(D|M_i)$ are the prior probability and evidence for model $M_i$, respectively. The quantity $\mathcal{P}(M_i|D)$ is precisely what we want to estimate; that is, the probability of the model, given the observed data. Most of the times, we assume $\Pi(M_1)=\Pi(M_0)=0.5$, as most DE models are purely phenomenological, and we cannot favour one over the other, purely on theoretical grounds. The ratio of Bayesian evidences should be weighted according to the reliability of the model priors. In particular, having values of $\Lambda$ orders of magnitude below quantum gravity expectations should heavily penalize \lcdm when compared against GREA. However, we have not taken this into account in our analysis and conclusions. We leave this analysis for a future publication.

Our analysis suggests that GREA can account for the data to some extent without invoking a fine-tuned (infinitesimally small and positive) cosmological constant. The best GREA model that fits the observational data point towards a {\em transient} phantom crossing at $z\leq 2$ (see \cref{fig:ev_w_rho}). 
We note that we have worked under the assumption of a \lcdm-expansion history at early times (in particular, at matter-radiation equality).  It would be interesting to explore how alternative initial conditions, e.g., with an Early Dark Energy (EDE)$-$like component or modified physics at recombination, can alter the derived constraints. Moreover, we have assumed that the main effect of GREA comes from the {\em homogeneous} entropy growth of the cosmic horizon. If GREA arises also from the gravitational collapse of {\em inhomogeneous} structures that form the cosmic web~\cite{Garcia-Bellido:2024tip}, it may modify the late universe dynamics in a way that could be tested with future cosmological data. This analysis, however, requires non-linear N-body simulations which go beyond the state-of-the-art. We leave this for future work.

Additional observational data will be critical to further assess the viability of GREA. Data at $z<1$, probing the growth of cosmic structures and the evolution of matter perturbations will be particularly interesting (\cref{fig:growth}). Forthcoming measurements from DESI and other stage-IV experiments will provide an opportunity to rigorously test the predictions of GREA across a wider range of cosmological observables.

\acknowledgments

R.C. would like to thank Vivian Poulin and Tristan L. Smith for insightful discussions. This work was supported by the
high-performance computing cluster Seondeok at the Korea Astronomy and Space Science Institute. 
R.C. is funded by the Czech Ministry of Education, Youth and Sports (MEYS) and European Structural and Investment Funds (ESIF) under project number CZ.02.01.01/00/22\_008/0004632.
J.G-B acknowledges support from the Research Project PID2021-123012NB-C43 [MICINN-FEDER], and the Centro de Excelencia Severo Ochoa Program CEX2020-001007-S at Instituto de F\'isica Te\'orica.
B.V.G would like to acknowledge the support from the European Union (ERC StG, LSS BeyondAverage, 101075919) and the Comunidad de Madrid 2019-T1/TIC-12702 grant. 
V.G.P. acknowledges support from Comunidad de Madrid in Spain (Atracci\'{o}n de Talento Contract grants: 2019-T1/TIC-12702 and 2023-5A/TIC-28943) and the Ministerio de Ciencia e Innovaci\'{o}n (MICINN, grant PID2021-122603NB-C21). 
This material is based upon work supported by the U.S. Department of Energy (DOE), Office of Science, Office of High-Energy Physics, under Contract No. DE–AC02–05CH11231, and by the National Energy Research Scientific Computing Center, a DOE Office of Science User Facility under the same contract. Additional support for DESI was provided by the U.S. National Science Foundation (NSF), Division of Astronomical Sciences under Contract No. AST-0950945 to the NSF’s National Optical-Infrared Astronomy Research Laboratory; the Science and Technology Facilities Council of the United Kingdom; the Gordon and Betty Moore Foundation; the Heising-Simons Foundation; the French Alternative Energies and Atomic Energy Commission (CEA); the National Council of Humanities, Science and Technology of Mexico (CONAHCYT); the Ministry of Science, Innovation and Universities of Spain (MICIU/AEI/10.13039/501100011033), and by the DESI Member Institutions: \url{https://www.desi.lbl.gov/collaborating-institutions}. Any opinions, findings, and conclusions or recommendations expressed in this material are those of the author(s) and do not necessarily reflect the views of the U. S. National Science Foundation, the U. S. Department of Energy, or any of the listed funding agencies.

The authors are honored to be permitted to conduct scientific research on I'oligam Du'ag (Kitt Peak), a mountain with particular significance to the Tohono O’odham Nation.

\section*{Data Availability}

Data from the plots in this paper will be made available on Zenodo (\url{https://zenodo.org/records/XXXXXX}) as part of DESI's Data Management Plan.

\appendix

\section{Parameter Constraints}
In \cref{tab:constraints}, we report the marginalized constraints on the cosmological parameters of GREA, for the various data combinations considered in this work.
\begin{table*}[t]
\caption{
    68\% credible intervals for the cosmological parameters in GREA, using various dataset combinations. We also report the \textit{maximum a posteriori} (MAP) values in parentheses, obtained with the minimizer \textsc{iMinuit} \cite{iminuit}. 
    \vspace{0.5em}}
    \label{tab:parameter_table2}
\centering
\small
\resizebox{0.95\textwidth}{!}{\begin{tabular}{ccccc}
\toprule
\toprule
Datasets & $\Omo$ & $10^2\omega_b$ & $H_0 [\kmsMpc]$ & $\alpha$  \\
\midrule[1.5pt] 
    DESI $\&$ Pantheon\texttt{+} & $0.323\pm 0.012$ & $-$ & $-$ & $1.1067^{+0.0044}_{-0.0049}$ \\  
    (Uncalibrated)   & $(0.316)$ & $(-)$ & $(-)$ & $(1.092)$\\  
   \midrule[1.5pt] 
   DESI $\&$ BBN & $0.300\pm 0.011$ & $2.23\pm 0.05$ & $70.9\pm 1.5$ & $1.105\pm 0.013$ \\
     & $(0.298)$ & $(2.23)$ & $(70.8)$ & $(1.104)$\\  
     \midrule  
    DESI $\&$ Union3 $\&$ BBN & $0.315\pm 0.010$ & $2.23\pm 0.05$ & $69.5\pm 1.3$ & $1.094\pm 0.012$ \\  
         & $(0.315)$ & $(2.23)$ & $(69.4)$ & $(1.092)$\\  
    \midrule  
     DESI \& DES-SN5YR $\&$ BBN & $0.326\pm 0.008$ & $2.23\pm 0.05$ & $68.6\pm 1.2$ & $1.088\pm 0.012$ \\ 
     & $(0.325)$ & $(2.23)$ & $(69.5)$ & $(1.087)$\\
    \midrule
     DESI $\&$ \panp~$\&$ BBN & $0.317\pm 0.009$ & $2.23\pm 0.05$ & $69.4\pm 1.3$ & $1.093\pm 0.012$ \\ 
     & $(0.316)$ & $(2.23)$ & $(69.4)$ & $(1.093)$\\
\midrule[1.5pt] 
     DESI $\&$ PR4 & $0.288\pm 0.004$ & $2.217\pm 0.014$ & $70.4\pm 0.5$ & $1.0994\pm 0.0029$ \\ 
     & $(0.287)$ & $(2.215)$ & $(70.4)$ & $(1.099)$\\
\midrule
DESI $\&$ P-ACT & $0.287\pm 0.004$ & $2.247\pm 0.011$ & $70.4\pm 0.5$ & $1.0962\pm 0.0023$ \\
& $(0.287)$ & $(2.247)$ & $(70.4)$ & $(1.096)$\\
\midrule

      DESI $\&$ \panp\ $\&$ PR4 & $0.292\pm 0.004$ & $2.216\pm 0.014$ & $69.9\pm 0.5$ & $1.0995\pm 0.0027$ \\ 
      & $(0.292)$ & $(2.215)$ & $(69.9)$ & $(1.099)$\\

\midrule

      DESI $\&$ \panp\ $\&$ P-ACT & $0.291\pm 0.004$ & $2.247\pm 0.011$ & $69.9\pm 0.5$ & $1.0961\pm 0.0024$ \\
      & $(0.291)$ & $(2.246)$ & $(69.8)$ & $(1.096)$\\
\midrule[1.5pt]
      DESI $\&$ \panp\ $\&$ \shoes & $0.317\pm 0.009$ & $2.68\pm 0.14$ & $72.9\pm 1.3$ & $1.092\pm 0.012$ \\ 
    $ $ & $(0.316)$ & $(2.646)$ & $(72.8)$ & $(1.093)$ \\
    \midrule
    DESI $\&$ \panp\ $\&$ PR4 $\&$ \shoes & $0.287\pm 0.004$ & $2.218\pm 0.014$ & $70.6\pm 0.5$ & $1.0997\pm 0.0029$ \\ 
      $ $ & $(0.287)$ & $(2.218)$ & $(70.6)$ & $(1.100)$ \\
    \midrule
    DESI $\&$ \panp\ $\&$ P-ACT $\&$ \shoes & $0.286\pm 0.004$ & $2.248\pm 0.011$ & $70.5\pm 0.5$ & $1.0967\pm 0.0024$ \\ 
      $ $ & $(0.285)$ & $(2.247)$ & $(70.6)$ & $(1.096)$ \\
\midrule
\midrule
\end{tabular}}\label{tab:constraints}
\end{table*}

\section{Comparison with DESI DR1}
\begin{figure}
    \centering
    \includegraphics[width=\linewidth]{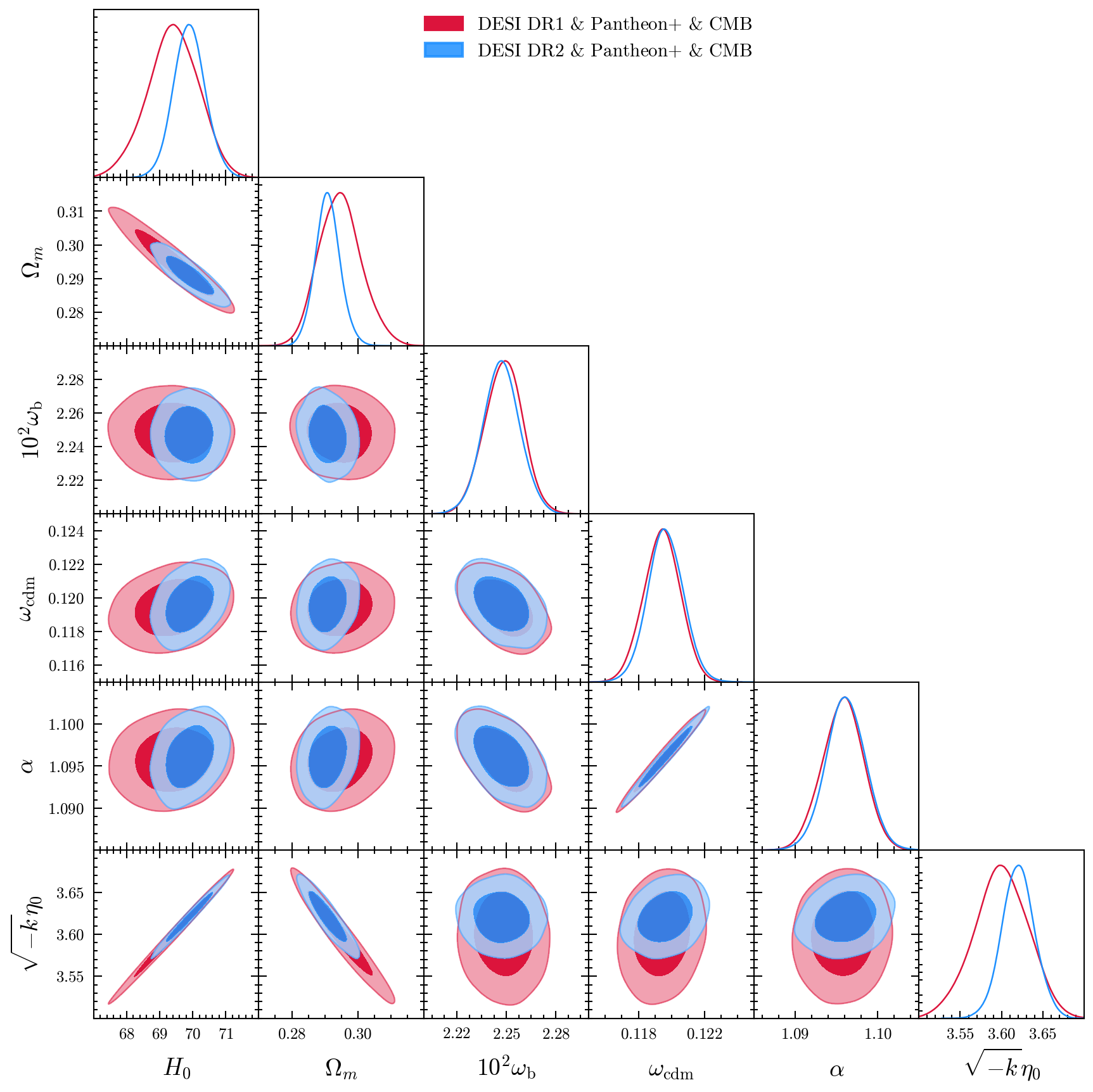}
    \caption{Similar to \cref{fig:DESI+BBN+SNe} but comparing the results using either DESI DR1 or DR2, combined with the \panp~  sample and the CMB data from P-ACT. The constraining power in the curvature scale parameter, $\sqrt{-k}\eta_0$, is improved by $60\%$
    when moving from DESI DR1 to DESI DR2, including also SNe Ia and CMB data.}
    \label{fig:full_posterior_dr1-vs-dr2}
\end{figure}

Throughout this work, we have used the BAO constraints from DESI DR2. We explore how the constrains have improved going from DR1 to DR2 in \cref{fig:full_posterior_dr1-vs-dr2}. DESI DR2 contains $6671$ dark tiles, increasing by a factor of $2.4$ the number of objects released in DR1\citep{DESI:2025zgx}.

We find that the constrains improved by a factor of $\sim2$ when moving from DR1 to DR2. The central values stay within $1\sigma$. The trends seen for the central values are similar to the results shown in \cref{fig:DESI+BBN+SNe} when removing the SN data from the DESI DR2 \& \panp~\&  CMB. In terms of the curvature scale parameter, $\sqrt{-k}\eta_0$, the constraints in DR2 are $60\%$ tighter than in DR1.

\section{CMB Compression and comparison with CamSpec PR4}
\begin{figure}
    \centering
    \includegraphics[width=\linewidth]{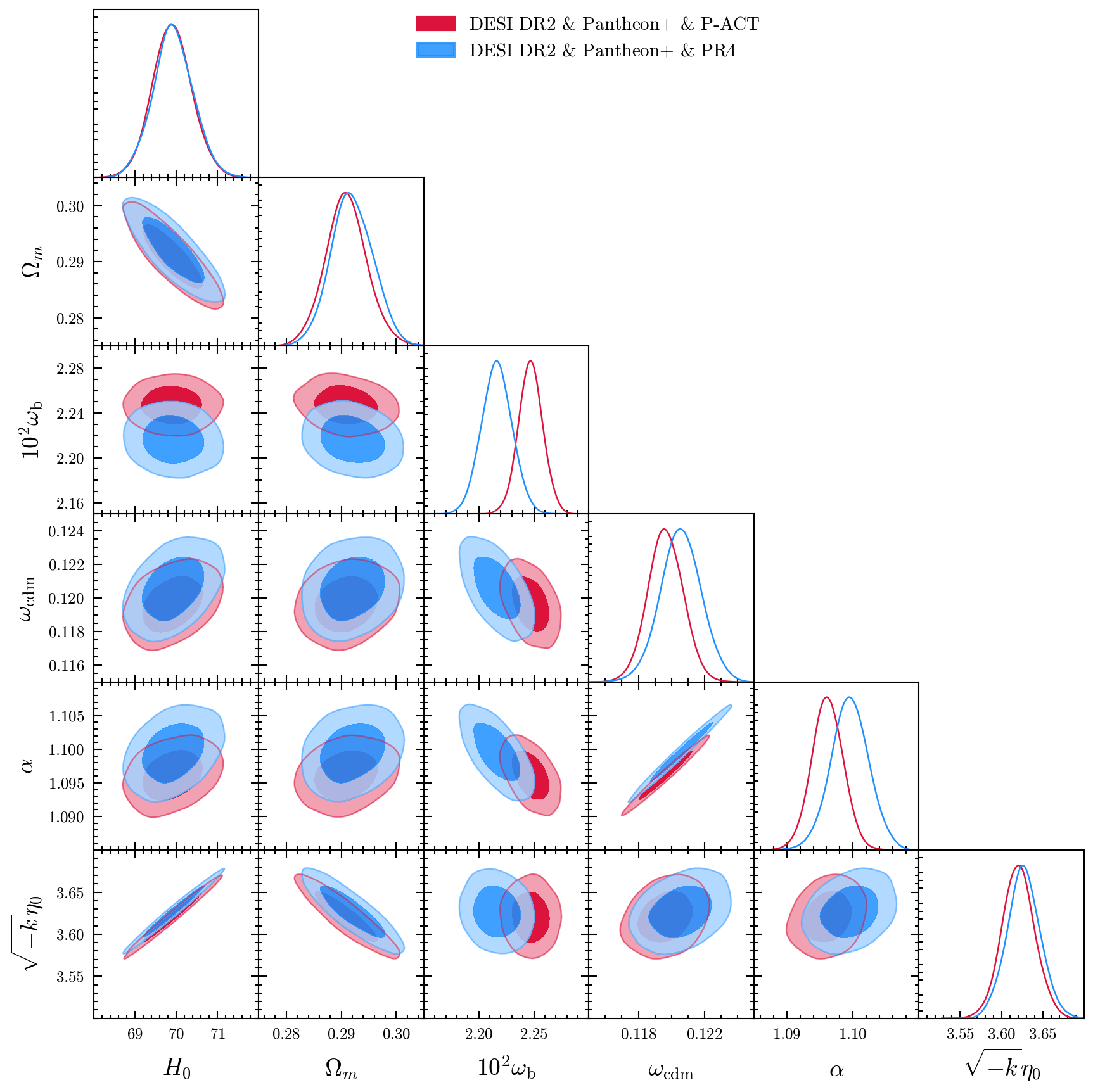}
    \caption{Comparison of the constraints obtained using the compressed CMB measurements of $(\theta_s,\omega_{\rm b},\omega_{\rm bc})$ extracted from either Planck (CamSpec) PR4, or the Planck PR3+ACT (P-ACT) combination.}
    \label{fig:full_posterior_pr4-vs-pact}
\end{figure}

In this appendix, we compare the constraints obtained using the CamSpec CMB likelihood, based on the Planck PR4 maps, with those obtained from the Planck and ACT (P-ACT) combination \cite{ACT:2025fju,ACT:2025tim}. These results are shown in\cref{fig:full_posterior_pr4-vs-pact}. The compressed CMB information from the Planck (PR4) chains can be found in Appendix A of Ref.~\cite{DESI:2025zgx}. Here, we provide the equivalent from the P-ACT chains \cite{ACT:2025fju}. The data vector and covariance matrix are given by

\begin{equation}
\vec{\mathcal{D}}=
\begin{pmatrix}
100\,\theta_s\\  \omega_{\rm b} \\ \omega_{\rm bc}
\end{pmatrix} =
\begin{pmatrix}
1.04165\\  0.0225046 \\ 0.141640
\end{pmatrix}
~,
\end{equation}

\begin{equation}
\Sigma=
\begin{pmatrix}
6.64717 \times 10^{-8} & 2.71701 \times 10^{-9} & -3.92400 \times 10^{-8} \\
2.71701 \times 10^{-9} & 1.20267 \times 10^{-8} & -3.49601 \times 10^{-8} \\
-3.92400 \times 10^{-8} & -3.49601 \times 10^{-8} & 1.11695 \times 10^{-6}
\end{pmatrix}
~,
\end{equation}
from which we construct the likelihood $-2\ln{\mathcal{L}}=\chi^2$ in the usual way, 
\begin{equation}
    \chi^2=\vec{r~}^{T}\cdot\Sigma^{-1}\cdot\vec{r}
\end{equation}
where $\vec{r}$ is the residual vector between the theoretical predictions and the data vector $\vec{\mathcal{D}}$.


\section{Author Affiliations}
\label{sec:affiliations}

\noindent \hangindent=.5cm $^{a}${CEICO, Institute of Physics of the Czech Academy of Sciences, Na Slovance 1999/2, 182 21, Prague, Czech Republic.}

\noindent \hangindent=.5cm $^{b}${Instituto de F\'{\i}sica Te\'{o}rica (IFT) UAM/CSIC, Universidad Aut\'{o}noma de Madrid, Cantoblanco, E-28049, Madrid, Spain}

\noindent \hangindent=.5cm $^{c}${Centro de Investigaci\'{o}n Avanzada en F\'{\i}sica Fundamental (CIAFF), Facultad de Ciencias, Universidad Aut\'{o}noma de Madrid, ES-28049 Madrid, Spain}

\noindent \hangindent=.5cm $^{d}${Korea Astronomy and Space Science Institute, 776, Daedeokdae-ro, Yuseong-gu, Daejeon 34055, Republic of Korea}

\noindent \hangindent=.5cm $^{e}${University of Science and Technology, 217 Gajeong-ro, Yuseong-gu, Daejeon 34113, Republic of Korea}

\noindent \hangindent=.5cm $^{f}${Lawrence Berkeley National Laboratory, 1 Cyclotron Road, Berkeley, CA 94720, USA}

\noindent \hangindent=.5cm $^{g}${Department of Physics, Boston University, 590 Commonwealth Avenue, Boston, MA 02215 USA}

\noindent \hangindent=.5cm $^{h}${Dipartimento di Fisica ``Aldo Pontremoli'', Universit\`a degli Studi di Milano, Via Celoria 16, I-20133 Milano, Italy}

\noindent \hangindent=.5cm $^{i}${INAF-Osservatorio Astronomico di Brera, Via Brera 28, 20122 Milano, Italy}

\noindent \hangindent=.5cm $^{j}${Department of Physics \& Astronomy, University College London, Gower Street, London, WC1E 6BT, UK}

\noindent \hangindent=.5cm $^{k}${Instituto de F\'{\i}sica, Universidad Nacional Aut\'{o}noma de M\'{e}xico,  Circuito de la Investigaci\'{o}n Cient\'{\i}fica, Ciudad Universitaria, Cd. de M\'{e}xico  C.~P.~04510,  M\'{e}xico}

\noindent \hangindent=.5cm $^{l}${Departamento de F\'isica, Universidad de los Andes, Cra. 1 No. 18A-10, Edificio Ip, CP 111711, Bogot\'a, Colombia}

\noindent \hangindent=.5cm $^{m}${Observatorio Astron\'omico, Universidad de los Andes, Cra. 1 No. 18A-10, Edificio H, CP 111711 Bogot\'a, Colombia}

\noindent \hangindent=.5cm $^{n}${Institut d'Estudis Espacials de Catalunya (IEEC), c/ Esteve Terradas 1, Edifici RDIT, Campus PMT-UPC, 08860 Castelldefels, Spain}

\noindent \hangindent=.5cm $^{o}${Institute of Cosmology and Gravitation, University of Portsmouth, Dennis Sciama Building, Portsmouth, PO1 3FX, UK}

\noindent \hangindent=.5cm $^{p}${Institute of Space Sciences, ICE-CSIC, Campus UAB, Carrer de Can Magrans s/n, 08913 Bellaterra, Barcelona, Spain}

\noindent \hangindent=.5cm $^{q}${University of Virginia, Department of Astronomy, Charlottesville, VA 22904, USA}

\noindent \hangindent=.5cm $^{r}${Fermi National Accelerator Laboratory, PO Box 500, Batavia, IL 60510, USA}

\noindent \hangindent=.5cm $^{s}${Center for Cosmology and AstroParticle Physics, The Ohio State University, 191 West Woodruff Avenue, Columbus, OH 43210, USA}

\noindent \hangindent=.5cm $^{t}${Department of Physics, The Ohio State University, 191 West Woodruff Avenue, Columbus, OH 43210, USA}

\noindent \hangindent=.5cm $^{u}${The Ohio State University, Columbus, 43210 OH, USA}

\noindent \hangindent=.5cm $^{v}${School of Mathematics and Physics, University of Queensland, Brisbane, QLD 4072, Australia}

\noindent \hangindent=.5cm $^{w}${Department of Physics, The University of Texas at Dallas, 800 W. Campbell Rd., Richardson, TX 75080, USA}

\noindent \hangindent=.5cm $^{x}${NSF NOIRLab, 950 N. Cherry Ave., Tucson, AZ 85719, USA}

\noindent \hangindent=.5cm $^{y}${Department of Physics, Southern Methodist University, 3215 Daniel Avenue, Dallas, TX 75275, USA}

\noindent \hangindent=.5cm $^{z}${Departament de F\'{i}sica, Serra H\'{u}nter, Universitat Aut\`{o}noma de Barcelona, 08193 Bellaterra (Barcelona), Spain}

\noindent \hangindent=.5cm $^{aa}${Institut de F\'{i}sica d’Altes Energies (IFAE), The Barcelona Institute of Science and Technology, Edifici Cn, Campus UAB, 08193, Bellaterra (Barcelona), Spain}

\noindent \hangindent=.5cm $^{ab}${Instituci\'{o} Catalana de Recerca i Estudis Avan\c{c}ats, Passeig de Llu\'{\i}s Companys, 23, 08010 Barcelona, Spain}

\noindent \hangindent=.5cm $^{ac}${Instituto de Astrof\'{i}sica de Andaluc\'{i}a (CSIC), Glorieta de la Astronom\'{i}a, s/n, E-18008 Granada, Spain}

\noindent \hangindent=.5cm $^{ad}${Departament de F\'isica, EEBE, Universitat Polit\`ecnica de Catalunya, c/Eduard Maristany 10, 08930 Barcelona, Spain}

\noindent \hangindent=.5cm $^{ae}${CIEMAT, Avenida Complutense 40, E-28040 Madrid, Spain}

\noindent \hangindent=.5cm $^{af}${Department of Physics, University of Michigan, 450 Church Street, Ann Arbor, MI 48109, USA}

\noindent \hangindent=.5cm $^{ag}${University of Michigan, 500 S. State Street, Ann Arbor, MI 48109, USA}

\noindent \hangindent=.5cm $^{ah}${National Astronomical Observatories, Chinese Academy of Sciences, A20 Datun Road, Chaoyang District, Beijing, 100101, P.~R.~China}

\bibliographystyle{JHEP}
\bibliography{biblio,DESI_supporting_papers}

\end{document}